\documentclass{ieeetmlcn}
\usepackage{fix-cm}
\usepackage{cite}
\usepackage{amsmath,amssymb,amsfonts}

\usepackage{graphicx,color}
\usepackage{textcomp}
\usepackage{xcolor}
\usepackage{csquotes}
\usepackage{bbm}
\usepackage{multirow}
\usepackage{hyperref}
\hypersetup{hidelinks}
\usepackage{algpseudocode}
\usepackage{algorithm}

\def\BibTeX{{\rm B\kern-.05em{\sc i\kern-.025em b}\kern-.08em
    T\kern-.1667em\lower.7ex\hbox{E}\kern-.125emX}}
\AtBeginDocument{\definecolor{tmlcncolor}{cmyk}{0.93,0.59,0.15,0.02}\definecolor{NavyBlue}{RGB}{0,86,125}}

\def\OJlogo{\vspace{-4pt}\includegraphics[height=18pt]{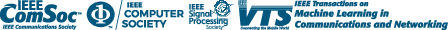}}
\def\seclogo{\vspace{10pt}\includegraphics[height=18pt]{images/new_logo.eps}}

\def\authorrefmark#1{\ensuremath{^{\textbf{#1}}}}

\begin{document}
\receiveddate{XX Month, XXXX}
\reviseddate{XX Month, XXXX}
\accepteddate{XX Month, XXXX}
\publisheddate{XX Month, XXXX}
\currentdate{XX Month, XXXX}
\doiinfo{TMLCN.2022.1234567}

\markboth{}{Mohammed M. H. Qazzaz {et al.}}

\title{xApp Empowered Resource Management for Non-Terrestrial Users in 5G O-RAN Networks}

\author{Mohammed M.H. Qazzaz\authorrefmark{1}, Syed Ali Zaidi\authorrefmark{1},\\ Aubida A. Al-Hameed\authorrefmark {2}, Abdelaziz Salama\authorrefmark{1}, and Des Mclernon\authorrefmark{1}
\affil{School of Electronic and Electrical Engineering, University of Leeds, Leeds, UK}
\affil{College of Electronics Engineering, Ninevah University, Mosul, Iraq}
\corresp{Corresponding author: Mohammed M.H. Qazzaz (email: ml14mmh@leeds.ac.uk).}
}


\begin{abstract}
This paper introduces a proactive Unmanned Aerial Vehicle (UAV) mobility management xApp for Open Radio Access Network (O-RAN) Near Real-Time Radio Intelligent Controller (Near-RT RIC) environments, employing Double Deep Q-Network (DDQN) reinforcement learning (RL) enhanced with transfer learning to optimise handover decisions for UAVs operating along predetermined flight trajectories. Unlike reactive approaches that respond to signal degradation, the proposed framework anticipates network conditions and minimises both outage probability and handover frequency through predictive optimisation. The system leverages centralised weight averaging to consolidate knowledge from multiple flight scenarios into a global model capable of generalising to previously unseen operational environments without extensive retraining. A comprehensive evaluation demonstrates that the proposed framework achieves a favourable trade-off between handover frequency and connectivity reliability, reducing handover events by up to 54.6\% compared to greedy approaches while maintaining outage probability at practically negligible levels. The results validate the effectiveness of intelligent learning-based approaches for UAV mobility management in next-generation O-RAN architectures, thereby contributing to seamless integration of aerial user equipment into cellular networks.
\end{abstract}

\begin{IEEEkeywords}
O-RAN, 5G, xApp, UAV Communication, Reinforcement Learning, RL, Double Deep Q-Network, DDQN, Resource Management, Handover Management, Non-Terrestrial Networks, NTN
\end{IEEEkeywords}

\maketitle

\section{INTRODUCTION}\label{ch3_introduction}
The proliferation of Unmanned Aerial Vehicles (UAVs), also known as \enquote{drones}, across civilian, commercial and military applications has necessitated their seamless integration into existing cellular communication infrastructure \cite{mishra2020survey}. Throughout this paper, the term \textit{non-terrestrial users} refers to aerial user equipment (UE) operating within the low-altitude airspace, as distinct from conventional ground-based terrestrial UE. While the 3GPP definition of Non-Terrestrial Networks (NTN) primarily encompasses satellite and High-Altitude Platform Station (HAPS) communication systems \cite{3gpp_ntn, 3gpp_ntn_2}, low-altitude UAVs share fundamental characteristics with NTN entities---including three-dimensional mobility, elevated line-of-sight propagation, and inter-cell interference patterns that differ markedly from terrestrial users. The mobility management solutions developed in this work for aerial UE address challenges that are foundational to the broader NTN integration roadmap, where seamless connectivity for diverse non-terrestrial platforms remains an open research objective. As 5G networks evolve toward more intelligent and adaptive architectures, the Open Radio Access Network (O-RAN) Alliance has introduced revolutionary concepts that enable artificial intelligence and machine learning (AI/ML) capabilities within network operations through intelligent Radio Intelligent Controllers (RICs) and extensible applications (rApps/xApps) \cite{wypior2022open,qazzaz2024non}. This convergence of UAV technology with intelligent cellular networks presents unprecedented opportunities for autonomous aerial operations while simultaneously introducing complex challenges in mobility and resource management \cite{mishra2020survey}.

UAV integration into cellular networks offers substantial advantages, including extended operational range, enhanced security compared to alternative connectivity solutions, and cost-effective utilisation of existing infrastructure \cite{mozaffari2019tutorial}. However, the unique characteristics of aerial mobility—including three-dimensional (3D) movement patterns, high line-of-sight (LoS) probabilities, and elevated altitudes—fundamentally differ from terrestrial user equipment, creating distinct challenges for traditional network management approaches. The dynamic nature of UAV operations, combined with stringent connectivity requirements for safety-critical applications, demands intelligent and proactive network management solutions that can anticipate and adapt to rapidly changing conditions \cite{dang2022deep}.

\subsection{Motivation}

Traditional cellular networks, designed primarily for terrestrial users, face significant challenges when serving UAVs as aerial user equipment. The high mobility and 3D movement of UAVs result in frequent handover events between base stations, potentially leading to service interruptions, increased signalling overhead, and radio link failures. These issues are exacerbated by the elevated altitude of UAV operations, which increases the likelihood of LoS interference and affects both aerial and terrestrial users \cite{new2021application}.

Effective handover management directly impacts radio resource allocation, spectrum efficiency, and bandwidth utilisation, making mobility optimisation inherently a resource management challenge. Conventional handover management schemes rely on reactive decision-making based on instantaneous signal measurements and predefined thresholds. Such approaches are inadequate for UAV scenarios where rapid mobility and predictable flight paths offer opportunities for proactive optimisation. The lack of intelligence in traditional handover mechanisms results in suboptimal resource utilisation and compromised quality of service for UAVs engaged in mission-critical operations \cite{meer2024mobility}.

The emergence of O-RAN architecture marks a paradigm shift toward intelligent, software-defined network management, facilitated by near real-time RICs equipped with AI/ML capabilities \cite{bonati2020open}. This architecture enables the deployment of intelligent xApps that can learn from network conditions and make autonomous decisions to optimise performance. However, existing mobility management solutions for UAVs have not fully leveraged the potential of O-RAN's intelligent capabilities, particularly in scenarios involving predetermined flight paths where predictive optimisation becomes feasible.

While traditional approaches assume reactive mobility management for unexpected user movement, many UAV applications operate along predetermined flight paths for mission planning, regulatory compliance, and safety considerations. These include surveillance missions, delivery routes and inspection tasks. Rather than viewing predetermined paths as a limitation, this presents an opportunity for proactive, predictive mobility optimisation that can anticipate network conditions and optimise handover decisions before connectivity issues arise. The flight paths used in this work are developed using our previously developed path planning approach that considers obstacles and basic connectivity constraints \cite{qazzaz2023low}. The assumption that the network has access to the UAV's predetermined flight path is well-grounded in current regulatory and architectural developments. In practice, many UAV operations — including delivery, surveillance, and inspection missions — require pre-filed flight plans as a regulatory obligation in numerous jurisdictions. The UTM system, described further in Section~\ref{ch3_xApp_impl}, is specifically designed to share such flight plan information with network infrastructure in real time. Furthermore, O-RAN specifications explicitly support the provision of UAV trajectory data as Enrichment Information (EI) to the Near-RT RIC \cite{polese2023understanding}, making flight path awareness a natural and increasingly standardised capability rather than an idealised assumption. Furthermore, the diverse nature of operational environments and flight patterns necessitates adaptive solutions that can generalise across different scenarios without requiring extensive retraining for each new deployment.

\subsection{Related Work}
Resource and mobility management remains a critical challenge in 5G/B5G networks, particularly with the emergence of non-terrestrial users and intelligent network architectures \cite{mishra2020survey}. The integration of AI/ML capabilities within O-RAN frameworks has opened new avenues for autonomous network optimisation. At the same time, UAV connectivity presents unique challenges that differ fundamentally from traditional terrestrial mobility patterns. Several studies have actively contributed to exploring and enhancing mobility and handover management in 5G and next-generation networks through various approaches and algorithms to address the challenges associated with efficient handover, seamless mobility, and optimised network performance.

The Open Radio Access Network Alliance has pioneered the integration of AI/ML applications through intelligent RICs, enabling near real-time network optimisation through extensible applications (xApps). Recent work by Orhan et al. \cite{orhan2021connection} developed a Graph Neural Network and reinforcement learning (RL) approach for connection management in O-RAN RIC, focusing on user-cell association and load balancing optimisation. However, their approach targets general user equipment without considering the unique mobility characteristics of aerial users. A recent study introduced a user handover–aware hierarchical federated learning framework for Open RAN-based next-generation mobile networks, where federated learning is employed to optimise decision-making across distributed RIC nodes while accounting for user mobility and handover dynamics \cite{singh2025user}. This approach highlights how hierarchical FL can address the scalability and privacy challenges of O-RAN while enhancing QoS in highly dynamic environments. The practical implementation challenges of DRL-based xApps in O-RAN environments have been explored by Seyfi et al. \cite{seyfi2025real}, who demonstrated RL-enabled xApps for experimental closed-loop optimisation using OSC RIC and srsRAN, highlighting both the feasibility and challenges of end-to-end AI-driven optimisation in disaggregated 5G frameworks.

Traditional handover management approaches, designed primarily for terrestrial users, face significant challenges when applied to UAV scenarios due to high mobility, 3D movement patterns, and elevated altitudes \cite{new2021application}. Wang et al. \cite{wang2018handover} employed a two-layer ML model using RL and deep neural networks to optimise handover controllers in large-scale heterogeneous networks, though without specific consideration for aerial mobility patterns. Recent advances in deep RL for UAV handover have shown promising results. Jang et al. \cite{jang2022proactive} presented a handover decision scheme using proximal policy optimisation to prevent unnecessary handovers while maintaining stable connectivity, achieving a reduction of up to 76\% in handover events compared to conventional schemes. However, their approach focuses on reactive decision-making rather than leveraging predictable flight paths for proactive optimisation. Kadir et al. \cite{kadir2024machine} provided a comprehensive review of ML-based approaches for handover decisions in cellular-connected UAVs, emphasising the potential of hybrid AI models combining deep RL with other techniques.

The integration of UAVs with other network entities for enhanced performance has gained attention in recent research. Oubbati et al. \cite{oubbati2025uav} proposed a UAV--Uncrewed Ground Vehicle (UGV) cooperative system, focusing on patrolling and energy management for urban monitoring, which demonstrates the potential for multi-agent coordination in aerial networks. While their work addresses energy efficiency in cooperative scenarios, it does not explicitly target cellular handover optimisation. Energy-efficient approaches to UAV communications have been explored through various perspectives. Dutriez et al. \cite{dutriez2024energy} developed an energy efficiency relaying election mechanism for 5G Internet of Things using deep reinforcement learning (DRL) techniques, providing insights into how RL can optimise energy consumption in wireless networks. Additionally, Alotaibi et al. \cite{alotaibi2025optimizing} investigated disaster response optimisation with UAV-mounted RIS and HAP-enabled edge computing in 6G networks, addressing large-scale dynamic environments and potential interference scenarios that complement our work's focus on robust connectivity management.

Previous research has addressed UAV path planning with connectivity considerations, though with fundamentally different objectives than our work. Xie et al. \cite{xie2021connectivity} proposed connectivity-aware 3D UAV path design using deep RL, focusing on end-to-end path optimisation from origin to destination, where both trajectory and connectivity are simultaneously optimised through a single learning framework. Similarly, Oh et al. \cite{oh2024deep} developed deep UAV path planning with assured connectivity in dense urban settings, emphasising real-time path adaptation based on connectivity constraints. These works represent comprehensive path planning solutions, whereas our approach leverages predetermined flight paths—common in many UAV applications for mission planning, regulatory compliance, and safety considerations—as an opportunity for proactive handover optimisation rather than viewing them as limitations. This work focuses specifically on advanced handover optimisation for UAVs operating along intelligently predetermined flight trajectories that consider obstacles and basic connectivity constraints.

\subsection{Contribution}
Enabling robust services to low-altitude UAVs using existing cellular networks presents significant challenges. High LoS probabilities in UAV-BS communication lead to interference issues, while the high speed and 3D movement of UAVs complicate handover management compared to ground UEs \cite{lin2019mobile}. Traditional reactive handover mechanisms fail to leverage the predictable nature of many UAV operations, leading to frequent handovers, radio link failures, and increased signalling overheads \cite{chen2020efficient}.

This work addresses these challenges by leveraging predetermined flight paths—common in surveillance, delivery, and inspection applications—as opportunities for proactive handover optimisation. While existing approaches either focus on general O-RAN optimisation or end-to-end path planning, our research targets explicitly handover optimisation for UAVs operating along intelligently predetermined trajectories. We present a DDQN-based xApp for O-RAN Near-RT RIC that anticipates network conditions and optimises handover decisions before connectivity degradation occurs. A key innovation is our transfer learning framework, which develops a global model from models trained across different flight trajectories and network conditions, significantly improving adaptability without requiring extensive retraining for each new deployment scenario.

The primary contributions include: (i) a proactive UAV mobility management xApp specifically designed for O-RAN architectures that transforms reactive handover decisions into predictive optimisation, (ii) a comprehensive transfer learning approach that generalises across different flight patterns and network conditions while reducing training overhead, and (iii) a thorough evaluation demonstrating that the proposed approach achieves a superior trade-off between handover frequency reduction and connectivity stability compared to traditional schemes, substantially reducing signalling overhead while maintaining outage probability at operationally acceptable levels.

\subsection{Structure of Paper}
This paper is structured as follows: Section \ref{ch3_introduction} introduces the motivation for this research, discusses related work, and outlines the contributions. Section II presents the fundamentals of O-RAN, including its architecture, disaggregation, and the ML framework within the O-RAN architecture. It also introduces RL and DDQN as enabling technologies for intelligent decision-making within the O-RAN framework. Section III details the system model, formulates the problem of UAV connectivity, and outlines the design of the proposed xApp. Section IV presents the evaluation of the proposed solution with performance results compared to the baseline schemes. Finally, Section V concludes the paper by summarising the key findings and contributions of this study.

\section{Background of Open Radio Access Networks (O-RAN)}

The Radio Access Network (RAN) serves as the critical interface between users and core network services, traditionally delivered as integrated hardware and software platforms from single vendors. The inception of the O-RAN Alliance marked a fundamental transformation toward disaggregated, intelligent, and interoperable network architectures. This paradigm shift enables the deployment of AI/ML-driven applications through programmable interfaces and intelligent controllers, directly supporting the advanced mobility management capabilities demonstrated in this work \cite{wypior2022open}.

\begin{figure}
  \includegraphics[width=\linewidth]{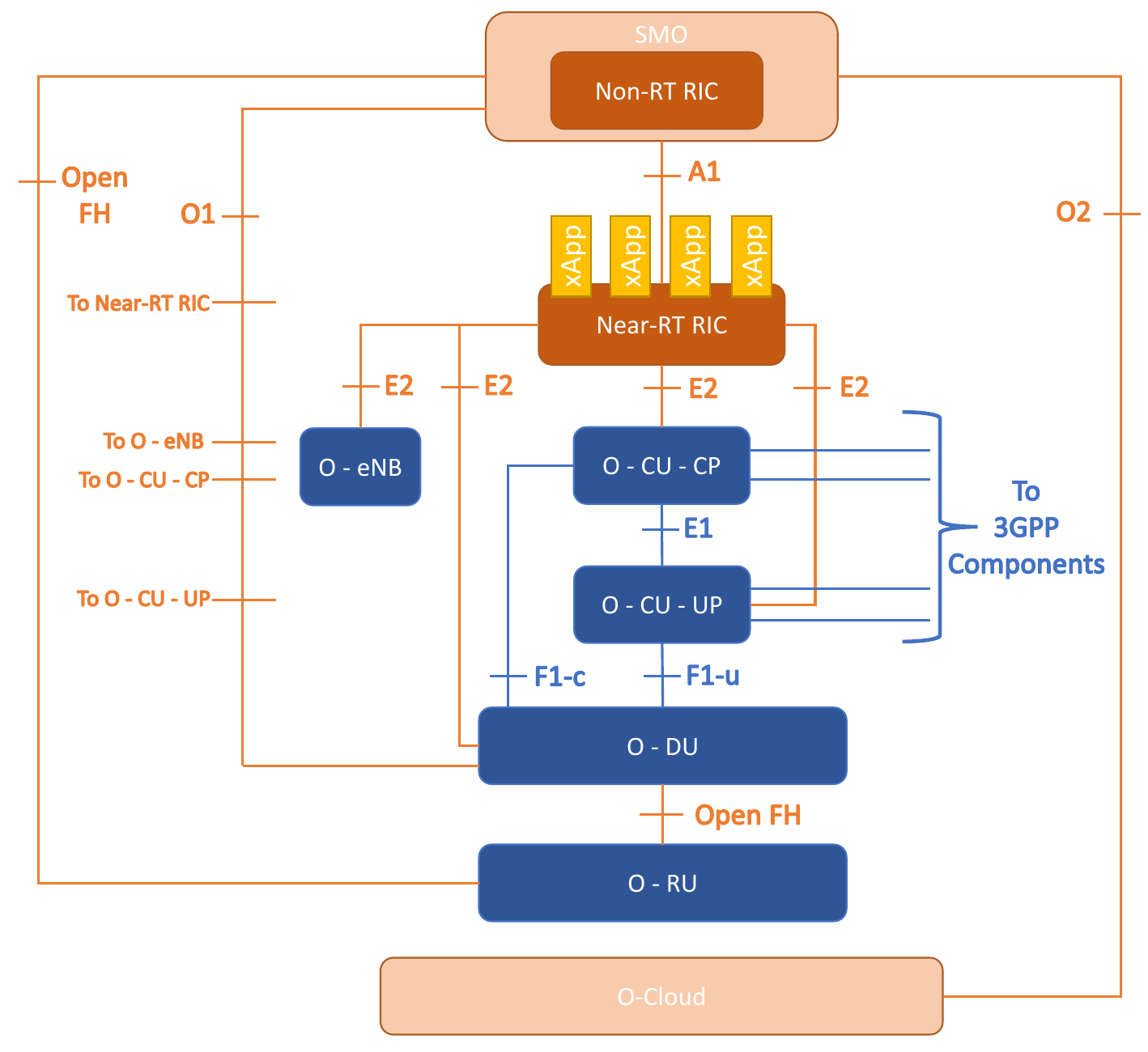}
  \caption{Open Radio Access Networks Architecture, with Components and Interfaces from O-RAN
and 3GPP \cite{alliance2020ran}}\label{ch3_ORAN_Arch}
\end{figure}

As illustrated in Figure \ref{ch3_ORAN_Arch}, the O-RAN architecture introduces fundamental disaggregation by splitting traditional base stations into distinct functional components. The Central Unit (CU) handles higher-layer protocols and is further divided into Control Plane (O-CU-CP) and User Plane (O-CU-UP) functions. The Distributed Unit (O-DU) manages MAC and RLC layer functions, while Radio Units (O-RU) handle the physical layer and RF functions. This disaggregation enables vendor-neutral interoperability through standardised open interfaces, promoting innovation and reducing deployment costs.

Central to O-RAN's intelligent capabilities are the RAN Intelligent Controllers (RICs), which operate at different time scales to optimise network performance. The Near Real-Time RIC (Near-RT RIC) provides sub-second control loops essential for mobility management applications, while the Non-RT RIC handles longer-term optimisation and policy decisions. The Near-RT RIC serves as the execution environment for intelligent applications called xApps, which can monitor network conditions, analyse performance metrics, and execute control decisions autonomously.

The E2 interface serves as the primary communication channel between the Near-RT RIC and RAN nodes (CUs and DUs), enabling both monitoring and control functions crucial for our UAV mobility management approach. Through E2 Service Models (E2SM), xApps can subscribe to Key Performance Metrics (KPMs) such as signal strength measurements, handover events, and user equipment reports. Simultaneously, the E2 interface supports RAN Control (E2SM-RC) functions, which enable xApps to influence handover parameters, trigger handover procedures, and adjust resource allocation decisions in real-time. The A1 interface connects the Non-RT RIC to the Near-RT RIC, facilitating policy updates and the distribution of ML models that enhance xApp performance over time.

The integration of AI/ML capabilities within O-RAN represents a fundamental shift from traditional rule-based network management to intelligent, adaptive optimisation. The O-RAN Alliance has established comprehensive guidelines for managing the ML model lifecycle within cellular networks, encompassing data collection from RAN nodes, model training and validation, deployment as xApps, and runtime inference and control. This framework enables the development of sophisticated learning algorithms that can adapt to dynamic network conditions and user behaviours, as demonstrated by our DDQN-based approach to UAV mobility management.

The xApp development model provides a standardised framework for deploying intelligent network functions within the Near-RT RIC. Each xApp operates as a microservice with defined interfaces for data collection, decision-making, and control execution. Through the E2 interface, xApps can access real-time network telemetry, including signal quality, user traffic patterns, and network topology information. This rich data environment, combined with the ability to execute control decisions with sub-second latency, creates an ideal platform for implementing proactive UAV mobility management strategies that anticipate network conditions and optimise handover decisions before connectivity degradation occurs.

The Service Management and Orchestration (SMO) framework manages the overall O-RAN deployment, including xApp lifecycle management, resource allocation, and integration with external systems such as Unmanned Aircraft System Traffic Management (UTM). For UAV applications, the SMO can receive flight-plan information via the EI interfaces, enabling xApps to incorporate predicted UAV trajectories into their optimisation algorithms. This architectural integration supports the proactive mobility management approach central to our work, in which predetermined flight paths serve as valuable input for anticipatory handover optimisation rather than as constraints on system flexibility.

The O-RAN framework embraces RL as a core enabler for intelligent network optimisation, where agents learn optimal policies through trial-and-error interactions with dynamic environments. Deep Q-Networks (DQN) and Double Deep Q-Networks (DDQN) extend traditional Q-learning using neural networks to approximate Q-functions in high-dimensional state spaces, making them particularly suitable for deployment as xApps within Near-RT RICs \cite{mnih2015humanlevel}. By combining experience replay and target network mechanisms, DDQN-based xApps can learn from network interactions, collecting feedback on signal quality, handover events, and performance metrics, to enable adaptive and proactive decision-making for tasks such as mobility management and resource allocation in O-RAN architectures \cite{xiong2019deep}.

\section{System Model}\label{ch3_Sys_Mod}
This study presents an intelligent xApp designed to optimise UAV mobility management within O-RAN architectures, specifically targeting handover decision-making for UAVs operating along predetermined flight trajectories. Unlike traditional reactive approaches that respond to connectivity degradation, our system leverages the predictable nature of UAV flight paths to enable proactive handover optimisation that anticipates network conditions and maintains seamless connectivity throughout the mission.

The proposed solution employs an RL-based DDQN framework deployed as an xApp within the O-RAN Near-RT RIC. Rather than addressing trajectory planning—which is handled through our previously developed obstacle-aware path generation approach \cite{qazzaz2023low}—this work focuses specifically on optimising radio resource management through intelligent handover decisions. The DDQN model learns optimal mobility policies by analysing real-time network conditions, UAV state information, and predicted flight path data to minimise both outage probability and handover frequency while ensuring stable connectivity.

A key innovation of our approach is the integration of the transfer learning concept that enables a global model to generalise across diverse flight scenarios without requiring extensive retraining for each new deployment. This capability addresses the practical challenge of deploying UAV mobility management solutions in varied operational environments while maintaining the benefits of learning-based optimisation.

\subsection{Problem Formulation}\label{ch.3_problem_form}

The primary objective of this work is to develop an xApp for proactive UAV mobility management within O-RAN environments. Unlike traditional reactive handover schemes that respond to signal degradation, our approach leverages predetermined flight trajectories to anticipate network conditions and optimise handover decisions before connectivity issues arise. This proactive strategy enables efficient radio resource management through reduced handover frequency and improved spectrum utilisation while maintaining reliable UAV connectivity throughout mission execution.

We consider a cellular network deployed over a coverage area $A = L \times W$, where $L$ and $W$ represent the geographical dimensions of the targeted region. The network infrastructure consists of $M$ terrestrial base stations $\mathcal{BS} = \{BS_1, BS_2, \ldots, BS_M\}$ with antenna heights denoted as $h_{BS}$. Each base station operates at carrier frequency $f_c$ and serves a heterogeneous set of user equipment, including both terrestrial users and aerial UAV clients.

\begin{figure*}
  \includegraphics[width=\linewidth]{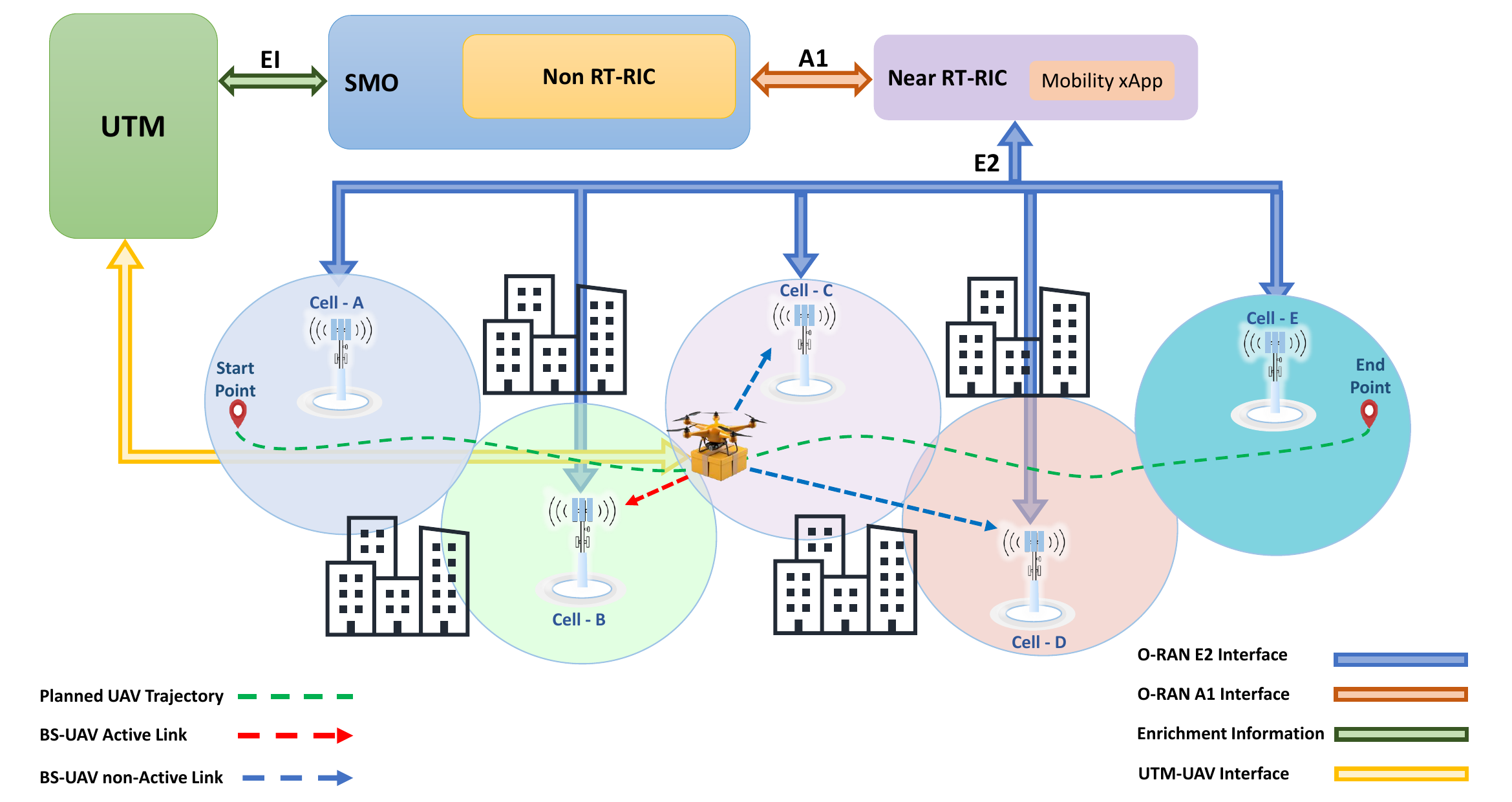}
  \caption{System Model} \label{ch3_system_model.fig}
\end{figure*}

The UAV operates along a predetermined 3D flight trajectory from initial position $\mathbf{p}_a = (x_a, y_a, h_a)$ to destination $\mathbf{p}_b = (x_b, y_b, h_b)$, where $h$ represents altitude above ground level. The UAV's position as a function of time is expressed as:
\[
\mathbf{p}_u(t) = (x_u(t), y_u(t), h_u(t)), \quad t \in [0, T]
\]
where $T$ denotes the total mission duration. At each moment, the UAV stays connected to a single base station selected through the proposed intelligent handover optimisation framework. Fig. \ref{ch3_system_model.fig} depicts this scenario, showcasing the UAV's trajectory and the distribution of ground BSs.

The optimisation framework targets two complementary objectives that directly impact radio resource efficiency: (i) \textbf{Outage Probability Minimisation}: Ensuring reliable signal quality throughout the flight by maintaining connection metrics above acceptable thresholds, thereby optimising spectrum utilisation and reducing interference to other network users; (ii) \textbf{Handover Frequency Reduction}: Minimising unnecessary handover events to reduce signaling overhead, preserve network resources, and eliminate service disruptions that compromise both UAV mission performance and overall network efficiency.

\subsubsection{Outage Probability}

The Signal-to-Interference-Plus-Noise Ratio (SINR) serves as the primary metric for assessing UAV communication link quality and determining optimal handover decisions in our framework. The SINR ($\gamma$) received by the UAV when connected to serving base station $k$ at time $t$ is expressed as \cite{tse2005fundamentals}:
\begin{equation}\label{gamma_eq}
\gamma_{k}(t)=\frac{h_{k}(t)P_{Tk}(t)}{N_{0}+\sum_{n\neq k}^{M}h_{n}(t)P_{Tn}(t)} ,
\end{equation}
where $k$ denotes the serving base station index, $M$ represents the total number of base stations in the network, $P_{Tk}$ is the transmit power of base station $k$, $h_k(t)$ represents the channel gain between the UAV and base station $k$ at time $t$, $N_0$ denotes the noise power density, and $h_n(t)$ and $P_{Tn}$ represent the channel gains and transmit powers of interfering base stations, respectively.

The channel model incorporates realistic LoS and non-line-of-sight (NLoS) propagation characteristics based on 3GPP specifications for UAV communications \cite{3gpp_36.777}. The probability of establishing a LoS link between the UAV and base station is given by:
\begin{equation}\label{case}
P_{\mathrm{LOS}}= 
\begin{cases} 
1, & \text{if } d_{2D} \leq d_1 \\ 
\frac{d_1}{d_{2D}} + \left(1 - \frac{d_1}{d_{2D}}\right)\exp\left(\frac{-d_{2D}}{p_1}\right) & \text{if } d_{2D} > d_1 
\end{cases}
\end{equation}
where $p_{1}=233.98\log_{10}(h_{ut}-0.95)$, $d_{2D}$ represents the horizontal distance between the UAV and base station, $h_{ut}$ denotes the UAV altitude (ranging from 22.5 to 100 meters), and $d_1=\max[(294.05\log_{10}(h_{BS})-432.94), 18]$ in meters, with $h_{BS}$ representing the base station antenna height.

The corresponding path loss models for LoS and NLoS conditions are:
\begin{equation}\label{los}
\begin{split}
    PL_{LOS} = \max\{((22.25
    - 0.5 \log_{10}(h_{ut}))\log_{10}(d_{3D}) \\ 
    + 30.9 + 20\log_{10}(f_c)),PL_{fr}\},
\end{split}
\end{equation}
\begin{equation}\label{nlos}
\begin{split}
    PL_{NLOS} = \max\{((43.2
    - 7.6 \log_{10}(h_{ut}))\log_{10}(d_{3D}) \\ 
    + 32.4 + 20\log_{10}(f_c)),PL_{LOS}\},
\end{split}
\end{equation}
where $d_{3D}$ is the three-dimensional distance between UAV and base station antennas, $f_c$ represents the carrier frequency in GHz, and $PL_{fr}$ denotes the free-space path loss:
\begin{equation}\label{pl}
    PL_{fr} = 20\log_{10}(d_{3D}) + 20\log_{10}(f_c) - 147.55 .
\end{equation}

An outage event occurs when the received SINR falls below the minimum threshold $\gamma_{th}$ required for reliable communication. The outage probability at time $t$ for UAV connection to base station $k$ is defined as:
\begin{equation}\label{eq:outage_prob}
    P_{out}(t) = \Pr(\gamma_{k}(t) < \gamma_{th}) .
\end{equation}

The optimisation objective seeks to minimise the average outage probability across the entire flight duration:
\begin{equation}
    \arg \min_{t}  \frac{1}{T} \sum_{t=1}^{T}P_{out}(t),
\end{equation}
where $T$ represents the total number of time steps. 
In practice, the outage probability is estimated empirically using a binary outage indicator function defined as:
\begin{equation}
    \mathcal{O}(t) = \mathbb{1}\left(\gamma_{k(t)}(t) < \gamma_{\text{th}}\right),
\end{equation}
where $\mathcal{O}(t) = 1$ indicates an outage event at time step $t$, and $\mathcal{O}(t) = 0$ otherwise. The empirical outage rate over the entire flight duration is then computed as:
\begin{equation}
    \hat{P}_{\text{out}} = \frac{1}{T} \sum_{t=1}^{T} \mathcal{O}(t),
\end{equation}
which serves as a Monte Carlo estimator of the theoretical outage probability defined in Eq.~(\ref{eq:outage_prob}). This empirical formulation is employed throughout the simulation and evaluation, where the outage rate is averaged across episodes to obtain statistically reliable performance metrics.

This can be formulated as a base station selection problem:
\begin{equation}\label{mini_SINR}
    \arg \min_{\{k(t)\}_{t=1}^{T}}  \frac{1}{T}\sum_{t=1}^{T}\Pr(\frac{h_{k}(t)P_{Tk}(t)}{N_{0}+\sum_{n\neq k(t)}^{M}h_{n}(t)P_{Tn}(t)} < \gamma_{th}) ,
\end{equation}
where $k(t)$ denotes the selected base station index at time $t$, determined through the proposed intelligent handover framework.

The channel gain $h_k(t)$ in Eq.~(\ref{mini_SINR}) is derived directly from the path loss models defined in Equations~(\ref{los})--(\ref{pl}). Specifically, the effective path loss between the UAV and base station $k$ at time $t$ is computed as the probabilistic combination of LoS and NLoS  components:
\begin{equation}
PL_k(t) = P_{\mathrm{LOS}} \cdot PL_{\mathrm{LOS}} + (1 - P_{\mathrm{LOS}}) \cdot PL_{\mathrm{NLOS}},
\end{equation}
where $P_{\mathrm{LOS}}$ is given by Eq.~(\ref{case}), and $PL_{\mathrm{LOS}}$, $PL_{\mathrm{NLOS}}$ are defined in Equations~(\ref{los}) and~(\ref{nlos}), respectively. The channel gain is then obtained as:
\begin{equation}
h_k(t) = 10^{-PL_k(t)/10},
\end{equation}
and equivalently for the interfering base stations $h_n(t)$, $\forall n \neq k$. This formulation ensures full consistency between the SINR expression in Eq.~(\ref{gamma_eq}) and the 3GPP-based channel model.

\subsubsection{Handover Probability}

To ensure connection stability and minimise unnecessary handovers, the proposed framework implements a handover strategy that prioritises maintaining connectivity with the current serving base station. At the same time, the received signal quality exceeds the minimum requirements. This approach reduces signalling overhead and prevents ping-pong effects that can degrade network performance. The handover decision logic is formulated through the following mathematical constraints:

\textbf{Connection Stability Condition:} The UAV maintains its current base station association when signal quality remains acceptable:
\begin{equation}
    x_k(t) = \mathbf{1}\left( x_k(t-1) = 1 \land \gamma_k(t) \geq \gamma_{\text{th}} \right),       
    \forall k,t,
\end{equation}
where $x_k(t)$ is a binary indicator variable equal to 1 if the UAV connects to base station $k$ at time $t$ and 0 otherwise, $\gamma_k(t)$ represents the SINR received from base station $k$ at time $t$, and $\gamma_{\text{th}}$ denotes the minimum SINR threshold for maintaining reliable communication. This constraint ensures that if the UAV was associated with base station $k$ at time $t-1$, the connection is preserved at time $t$ only if the received SINR remains above the threshold.

\begin{figure*}
  \includegraphics[width=\linewidth]{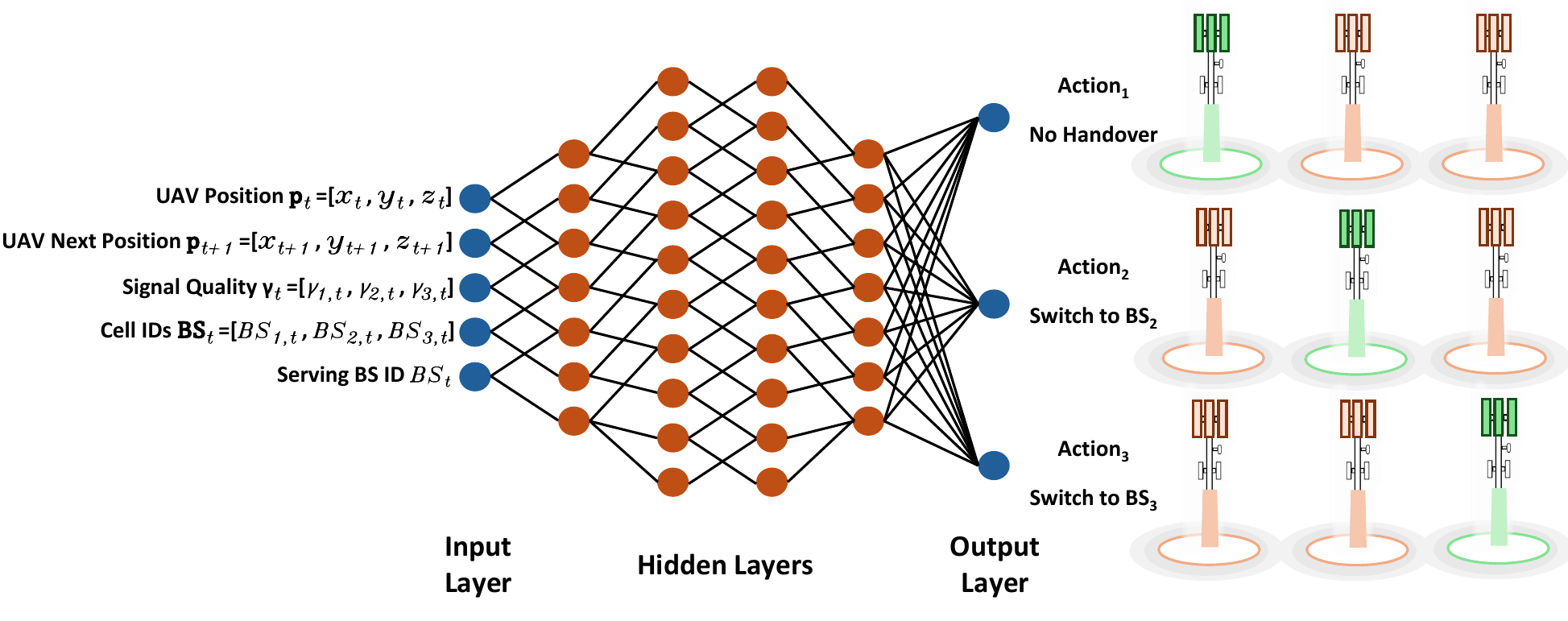}
  \caption{DDQN Model Layers Architecture}\label{ch3_DQN_fig}
\end{figure*}

\textbf{Handover Trigger Condition:} A mandatory handover occurs when the current serving base station's signal quality becomes insufficient:
\begin{equation}
\mathbf{1}\left( x_k(t-1) = 1 \ \land \ \gamma_k(t) < \gamma_{\text{th}} \right) 
= \sum_{\substack{j=1 \\ j \neq k}}^{M} x_j(t), \quad \forall k,t
\end{equation}
The left-hand side equals 1 when the UAV was connected to base station $k$ at time $t-1$ and the SINR from that base station drops below $\gamma_{\text{th}}$ at time $t$. The right-hand side enforces connection to exactly one alternative base station $j \neq k$, ensuring seamless handover execution. When signal quality remains adequate, the left-hand side equals 0, preventing unnecessary handover attempts.

\textbf{Proactive Handover Decision:} Beyond reactive handover triggers, the framework considers proactive handover opportunities when significantly better serving options become available:
\begin{equation}\label{HO_indicator}
HO(t) = \mathbf{1}\bigl\{\exists j \neq k, \gamma_j(t) > \gamma_k(t) + \gamma_{\text{mgn}} 
\text{ and } x_k(t-1) = 1\bigr\},
\end{equation}
where $\gamma_{\text{mgn}}$ represents the handover margin parameter that prevents frequent handovers due to minor signal fluctuations. A handover event occurs when two conditions are satisfied simultaneously: (i) an alternative base station $j$ provides SINR exceeding the current serving base station $k$ by at least the handover margin $\gamma_{\text{mgn}}$, and (ii) the UAV was connected to base station $k$ in the previous time step. This formulation balances connection stability with opportunities for improved service quality, enabling the learning algorithm to optimise long-term connectivity performance rather than responding solely to instantaneous signal measurements.

\subsection{Reinforcement Learning Model}

While the handover constraints defined in Section \ref{ch.3_problem_form} establishes the fundamental conditions for execution, which reflect reactive decision-making based on instantaneous signal measurements. Optimal timing of handover and target base station selection in dynamic UAV environments requires adaptive intelligence that accounts for multiple factors: predicted signal quality, interference from neighbouring cells, UAV trajectory, and historical network performance. Traditional rule-based approaches cannot learn from experience or adapt to varying conditions, leading to suboptimal resource utilisation and unnecessary handovers. The proposed RL framework enables proactive optimisation by learning policies that anticipate connectivity requirements based on UAV flight path information and network state evolution. Rather than reacting to signal threshold violations, the model knows to minimise outage probability and reduce handover frequency through predictive decision-making that considers the long-term impact of current actions.

\subsubsection{Double Deep Q-Network Approach}
The core learning algorithm utilises a DDQN to mitigate the overestimation bias inherent in traditional DQN. DDQN separates action selection and evaluation by using the online network for action selection and the target network for value estimation, yielding more stable learning in environments with noisy or sparse rewards. The DDQN update rule is:
\begin{equation} 
\begin{split} 
Q_{\theta}(s,a) \leftarrow Q_{\theta}(s,a) + \alpha \big[ \, r + \gamma Q_{\theta^-}(s', & \arg\max_{a'} Q_{\theta}(s', a')) \\ & - Q_{\theta}(s,a) \, \big] 
\end{split} 
\end{equation}

where $Q_{\theta}$ is the online network, $Q_{\theta^-}$ the target network, $\alpha$ the learning rate, and $\gamma$ the discount factor. This formulation reduces overoptimistic value estimates that hinder convergence. For comparison, we also evaluate the standard DQN, which uses a single network for both roles, to highlight the benefits of bias reduction in UAV handover optimisation.

Initial experiments explored policy gradient methods, including Proximal Policy Optimisation (PPO) and Actor-Critic (A2C) algorithms. While these methods are theoretically capable of handling sequential decision-making, our experiments revealed practical limitations in this context. Specifically, policy-gradient approaches tended to converge prematurely to near-deterministic “no-handover” policies. This outcome can be attributed to the high variance of gradient estimates and the strong influence of short-term penalties (handover costs) on the policy update, which biased the optimisation toward immediate reward maximisation and discouraged exploration of policies with long-term benefits. 

In contrast, the discrete and finite nature of the action space, coupled with the delayed-reward structure of the handover problem, makes value-based methods such as DQN and DDQN more suitable. These methods directly approximate action-value functions, enabling more effective propagation of long-term returns and facilitating stable learning in environments where current decisions substantially influence future connectivity outcomes.

\begin{algorithm}
\caption{DDQN Training Algorithm for Outage- and Handover-Aware UAV Mobility Management}
\label{ch3_xApp_algo}
\begin{algorithmic}[1]
\State \textbf{Initialise:} Learning rate $\alpha$, discount factor $\gamma$, exploration rate $\epsilon$, batch size $B$, target network update interval $C$, exploration decay rate $\epsilon_{\text{decay}}$
\State \textbf{Initialise :} Environment parameters: UAV flight area, base station coordinates, SINR threshold $\gamma_{\text{th}}$, handover margin $\gamma_{\text{mgn}}$, number of training paths and episodes
\State \textbf{Initialise:} Weighting coefficients $\alpha_o$, $\beta_h$ for outage and handover penalties in the reward function
\State \textbf{Initialise:} DDQN with online network $Q_{\theta}$ and target network $Q_{\theta^{-}} \leftarrow Q_{\theta}$
\State \textbf{Initialise:} Replay memory buffer $\mathcal{D}$

\For{each training episode $m = 1, \dots, M$}
    \State Obtain initial state vector $S_t = \{\mathbf{p}_t, \mathbf{p}_{t+1}, \boldsymbol{\gamma}_t, \mathbf{BS}_t, BS_t\}$
    \State Set cumulative reward $R_{\text{episode}} \gets 0$, handover count $H_{\text{episode}} \gets 0$, outage count $O_{\text{episode}} \gets 0$
    \State Set episode done flag: $\text{done} \gets \text{False}$
    
    \While{not done}
        \State Select action $A_t$ using $\epsilon$-greedy policy:
        \[
        A_t = 
        \begin{cases}
        \text{random action}, & \text{with prob. } \epsilon \\
        \arg\max_a Q_{\theta}(S_t, a), & \text{otherwise}
        \end{cases}
        \]
        
        \State Execute action $A_t$, observe next state $S_{t+1}$, reward $r_t$, outage indicator $o_t$, and done flag
        \State Store transition $(S_t, A_t, r_t, S_{t+1}, \text{done})$ in buffer $\mathcal{D}$
        
        \State Sample random minibatch $\{(S_j, A_j, r_j, S_{j+1}, \text{done}_j)\}_{j=1}^{B}$ from $\mathcal{D}$
        \For{each sampled transition}
            \State Compute best action: $a^* = \arg\max_{a'} Q_{\theta}(S_{j+1}, a')$
            \State Compute target:
            \[
            y_j = 
            \begin{cases}
            r_j, & \text{if } \text{done}_j = \text{True} \\
            r_j + \gamma Q_{\theta^{-}}(S_{j+1}, a^*), & \text{otherwise}
            \end{cases}
            \]
            \State Update $Q_{\theta}$ by minimizing loss: $L = \frac{1}{B} \sum_j (y_j - Q_{\theta}(S_j, A_j))^2$
        \EndFor
        
        \If{current step is divisible by $C$}
            \State Update target network: $\theta^{-} \gets \theta$
        \EndIf
        
        \State Update current state: $S_t \gets S_{t+1}$
        \State Update counters: $R_{\text{episode}} \mathrel{+}= r_t$, $H_{\text{episode}} \mathrel{+}= \text{HO}(t)$, $O_{\text{episode}} \mathrel{+}= \mathcal{O}(t)$
    \EndWhile
    \State Optionally: Decay exploration rate $\epsilon \gets \epsilon \cdot \epsilon_{\text{decay}}$
\EndFor
\end{algorithmic}
\end{algorithm}

\subsubsection{State Space, Action Space, and Reward Design}
The DDQN agent operates within a carefully designed state-action framework that captures the essential elements of UAV mobility and network conditions. The state space includes spatial information (current and following UAV positions), signal quality measurements (SINR from the three most relevant base stations), and connectivity context (serving cell identification) as shown in Fig. \ref{ch3_DQN_fig}. This design provides sufficient information for intelligent handover decisions while maintaining computational tractability.
\textbf{State Space Definition:} Each state is characterised by factors that directly influence mobility decisions:
\begin{itemize}
  \item \textbf{UAV Position Vector:} Current 3D coordinates of the UAV at time $t$ $\mathbf{p}_t = [x_t, y_t, z_t]^T \in \mathbb{R}^3$
  \item \textbf{Next UAV Position Vector:} Next UAV position from flight path $\mathbf{p}_{t+1} = [x_{t+1}, y_{t+1}, z_{t+1}]^T \in \mathbb{R}^3$
  \item \textbf{Signal Quality Vector:} SINR measurements from three candidate base stations $\boldsymbol{\gamma}_t = [\gamma_{1,t}, \gamma_{2,t}, \gamma_{3,t}]^T \in \mathbb{R}^3$
  \item \textbf{Cell Identifier Vector:} Base station IDs $\mathbf{BS}_t = [BS_{1,t}, BS_{2,t}, BS_{3,t}]^T \in \mathbb{Z}^3$
  \item \textbf{Serving Cell Indicator:} Current connection index $BS_t \in \{BS_{1,t}, BS_{2,t}, BS_{3,t}\}$
\end{itemize}
The complete state representation is:
\begin{equation}\label{State}
S_t = \left\{ \mathbf{p}_t, \mathbf{p}_{t+1}, \boldsymbol{\gamma}_t, \mathbf{BS}_t, BS_t \right\}
\end{equation}
\textbf{Action Space Definition:} The agent selects from three discrete actions at each decision epoch:
\begin{equation}\label{Action}
A_t = \{a_1, a_2, a_3\}
\end{equation}
where $a_1$ maintains the current base station connection, $a_2$ switches to the second-strongest candidate base station, and $a_3$ switches to the third candidate base station. Each action directly determines the serving base station $k(t)$ for the subsequent state: action $a_1$ maintains the current association such that $k(t) = k(t-1)$, while actions $a_2$ and $a_3$ result in a handover to the respective candidate base station, yielding $k(t) \neq k(t-1)$.

\textbf{Reward Function Design:} The reward function is designed to evaluate the consequence of the agent's selected action at each decision step. Since the action $A_t$ determines the serving base station $k(t)$, both components of the reward---outage penalty and handover penalty---are implicitly conditioned on the selected action through the resulting base station association. This coupling ensures that the learning signal directly reflects the quality of the control decision.

To address the challenge of balancing continuous outage values with discrete handover events, the handover penalty employs a sigmoid-based smooth approximation. Given the action-determined serving base station $k(t)$, the handover penalty is formulated as:
\begin{equation}\label{soft_HO}
\begin{aligned}
\widetilde{\text{HO}}(t \mid A_t)
&= \mathbf{1}\left(k(t) \neq k(t-1)\right) \cdot \eta \\
&\quad \cdot \sigma\left(
\frac{\gamma_{k(t)}(t) - \gamma_{k(t-1)}(t) - \gamma_{\text{mgn}}}{\tau}
\right)
\end{aligned}
\end{equation}
where $k(t)$ is the base station selected as a result of action $A_t$, $k(t-1)$ is the previously serving base station, $\sigma(z) = \frac{1}{1 + e^{-z}}$ is the sigmoid function, $\tau$ controls transition sharpness, $\eta$ provides scaling alignment, and $\gamma_{\text{mgn}}$ represents the handover margin parameter. The indicator $\mathbf{1}(k(t) \neq k(t-1))$ ensures that the penalty is applied only when the agent's action results in an actual base station change, thereby directly linking the handover cost to the control decision.

Similarly, the outage component of the reward depends on the action through the resulting connection. The outage penalty at time $t$ is determined by the SINR received from the action-selected base station $k(t)$:
\begin{equation}\label{outage_reward}
P_{\text{out}}(t \mid A_t) = \mathcal{O}(t) = \mathbf{1}\left(\gamma_{k(t)}(t) < \gamma_{\text{th}}\right).
\end{equation}
The complete reward function balances both optimisation objectives:
\begin{equation}\label{reward}
R_t(A_t) = - \alpha_o \cdot P_{\text{out}}(t \mid A_t) - \beta_h \cdot \widetilde{\text{HO}}(t \mid A_t),
\end{equation}
where $\alpha_o$ and $\beta_h$ are weighting coefficients. For this study, equal weighting ($\alpha_o = \beta_h$) provides balanced optimisation of both outage minimisation and handover reduction. This selection allows a clear assessment of the fundamental learning behaviour without introducing complexities from fine-tuning these specific weights. A detailed analysis of varying $\alpha_o$ and $\beta_h$ to explore different operational trade-offs is reserved for future study.

Algorithm \ref{ch3_xApp_algo} presents the complete DDQN training procedure, which applies equally to both DDQN and standard DQN implementations with appropriate modifications to the Q-value update mechanism.

\subsection{Transfer Learning}

The challenge of developing generalisable UAV mobility management solutions across diverse flight scenarios necessitates an intelligent approach to knowledge reuse and model adaptation. Rather than training separate models for each possible flight path—an approach that risks overfitting and computational inefficiency—this study employs a transfer learning framework that consolidates knowledge from multiple independently trained models into a global model capable of generalising to previously unseen scenarios. The consolidation is performed through centralised weight averaging, inspired by the federated averaging principle introduced by McMahan et al.~\cite{mcmahan2017fedavg}, but adapted to suit the offline, non-distributed nature of the training pipeline in this work.

The fundamental premise underlying this approach is that DQN and DDQN models learn generalisable handover decision policies based on signal quality patterns, interference characteristics, and UAV-network spatial relationships rather than memorising specific coordinate sequences. The learned Q-function captures optimal handover timing based on SINR differentials, anticipated signal degradation patterns, and target base station selection criteria—knowledge that exceeds specific flight geometries while adapting to local network conditions. The state representation, incorporating signal measurements, spatial relationships, and following UAV positions, enables the model to develop decision policies that focus on fundamental UAV-network interaction patterns applicable across diverse operational scenarios.

\begin{algorithm}[t]
\caption{Centralised Weight Averaging and Transfer Learning for Global DDQN Model}
\label{ch_3_alg_transfer_learning}
\begin{algorithmic}[1]
\State \textbf{Input:} Set of pre-trained DDQN models $\{Q_{\theta_1}, Q_{\theta_2}, \dots, Q_{\theta_N}\}$, each trained on distinct UAV trajectory representing diverse operational scenarios
\State \textbf{Input:} Target flight path for new UAV deployment task
\State \textbf{Input:} Reduced learning rate $\alpha_{\text{fine}}$ for fine-tuning phase

\State \textbf{Phase 1: Centralised Weight Averaging}
\State \textbf{Initialise:} Global DDQN model $Q_{\theta_{\text{global}}}$
\State \textbf{Aggregate:} Compute average weights across pre-trained models:
\[
\theta_{\text{global}} \gets \frac{1}{N} \sum_{i=1}^{N} \theta_i
\]
\State \textbf{Assign:} Set $Q_{\theta_{\text{global}}}$ with aggregated weights

\State \textbf{Phase 2: Transfer Learning Adaptation}
\State \textbf{Freeze:} Lower-layer parameters $\theta_f$ of global model to preserve consolidated handover decision patterns
\State \textbf{Initialise:} Upper-layer trainable parameters $\theta_t$ for scenario-specific adaptation
\State \textbf{Fine-tune:} Train only $\theta_t$ on target flight path using reduced learning rate $\alpha_{\text{fine}}$ while minimizing:
\[
\mathcal{L}_t(f(x; \theta_f, \theta_t))
\]
\State \textbf{Monitor:} Convergence and prevent catastrophic forgetting of federated knowledge
\State \textbf{Output:} Fine-tuned global model $Q_{\theta_{\text{global}}^{*}}$ adapted for target UAV deployment scenario
\end{algorithmic}
\end{algorithm}

The federated averaging process consolidates knowledge from multiple individually trained DQN and DDQN models through weight aggregation:
\[
\theta_{\text{global}} = \frac{1}{N} \sum_{i=1}^{N} \theta_i
\]
where $\theta_{\text{global}}$ represents the parameters of the global model, $\theta_i$ are the parameters of the $i$-th individual model trained on a specific flight path, and $N$ is the total number of models. This averaging process effectively combines the learned handover policies from diverse training scenarios, creating a consolidated knowledge base that captures common patterns while avoiding overfitting to any single flight path.

It is important to note that the weight aggregation procedure employed in this work differs from standard federated learning (FL) protocols in several key aspects. In conventional FL, multiple distributed clients train on local data and iteratively communicate model updates to a central server over multiple communication rounds, with the primary motivation being data privacy preservation and communication efficiency \cite{mcmahan2017fedavg}. In contrast, our approach performs a single-round centralised weight averaging over models that have been fully trained offline on distinct flight path scenarios, without privacy constraints or iterative client-server communication. This design choice is motivated by the practical characteristics of UAV mobility management: training data for different flight paths is generated through simulation and can be centrally collected, eliminating the need for iterative distributed protocols. The term \textit{weight averaging} is therefore used throughout to describe this consolidation process, which draws conceptual inspiration from federated averaging while operating in a centralised offline training pipeline.

The individual models are trained on carefully selected flight paths that represent diverse operational scenarios, including varying altitudes (20-30m), different trajectory patterns, and network topologies. This diversity ensures that the aggregated knowledge captures representative UAV mobility patterns and handover decision contexts rather than path-specific geometries. The selection strategy emphasises coverage of different signal environments, variations in base station density, and interference conditions that UAVs typically encounter in operational deployments.

Following federated averaging, the global model undergoes transfer learning adaptation for new flight scenarios through selective fine-tuning:
\[
\theta_t^* = \arg \min_{\theta_t} \mathcal{L}_t(f(x; \theta_f, \theta_t))
\]
where $\mathcal{L}_t$ denotes the loss function for the target flight path, $\theta_f$ represents frozen parameters that preserve the consolidated federated knowledge, and $\theta_t$ denotes trainable parameters adapted to the new scenario. This approach maintains the generalisable handover decision patterns while allowing adaptation to specific network conditions and flight characteristics.

The fine-tuning process employs layer-wise parameter freezing, where early layers, which capture fundamental UAV-network interaction patterns, remain fixed, while deeper layers adapt to scenario-specific nuances. This strategy prevents catastrophic forgetting of valuable consolidated knowledge while enabling efficient adaptation to new operational conditions with reduced learning requirements and improved convergence stability.

While this approach cannot guarantee optimal performance across all conceivable flight scenarios, it demonstrates effective generalisation within tested operational conditions representing diverse UAV applications. The method's effectiveness depends on the representativeness of training scenarios and the extent to which fundamental handover optimisation principles apply across different operational contexts. Comprehensive validation through testing on previously unseen flight paths, including detailed performance metrics and weight similarity analysis, is presented in Section \ref{ch3_results} to demonstrate the practical effectiveness of this transfer learning framework. Algorithm \ref{ch_3_alg_transfer_learning} outlines the complete weight averaging and transfer learning procedure, explaining how centralised knowledge consolidation and selective adaptation enable efficient deployment of UAV mobility management solutions across diverse operational scenarios.

\begin{figure*}[t]
  \includegraphics[width=\linewidth]{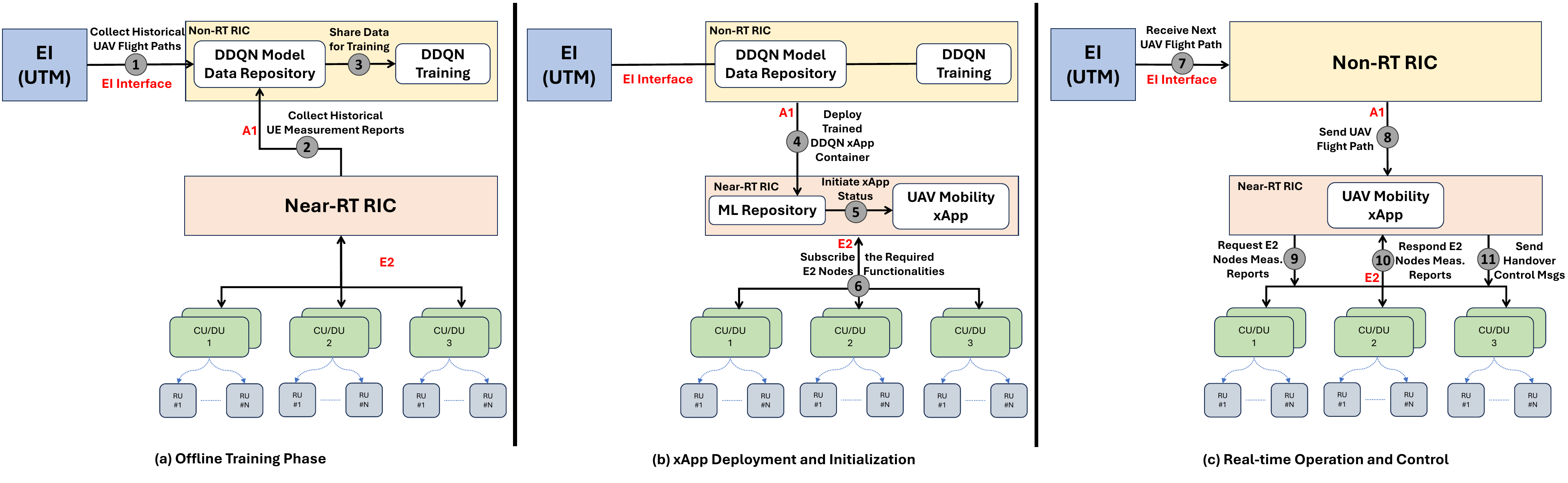}
  \caption{xApp workflow within the O-RAN Framework} \label{ch3_xApp_cycle.fig}
\end{figure*}

\subsection{xApp Implementation in O-RAN Architecture}\label{ch3_xApp_impl}
In O-RAN Near-RT-RIC, the xApps are designed as microservices that receive input data through interfaces between the O-RAN RIC and RAN functionality, and provide additional functionality as output to the RAN. Our UAV Mobility Management xApp is designed to ensure seamless connectivity and optimised network resources for UAVs in O-RAN networks. This xApp operates within the Near-RT-RIC and consists of several key components that work together to achieve intelligent mobility management. The Near-RT RIC environment offers sub-second control loops, which align well with UAV handover timing requirements, where decision latencies of hundreds of milliseconds are acceptable for UAVs operating at typical speeds and altitudes. A block diagram illustrating the xApp implementation and its lifecycle is shown in Fig. \ref{ch3_xApp_cycle.fig}. This diagram illustrates the flow of information between the Near-RT-RIC and surrounding entities, including the Non-RT-RIC, the EI, and the O-RAN E2 nodes (CU/DU), down to the aerial user equipment (the UAV).

Unmanned Aircraft System Traffic Management (UTM) is a system being developed to manage low-altitude drone traffic, particularly for Beyond-Visual-Line-of-Sight (BVLOS) operations \cite{kopardekar2016unmanned}. It aims to facilitate safe and efficient drone integration into airspace by providing functionalities like airspace organisation, traffic management, and real-time tracking. For this model, we will utilise UTM data as EI to inform the xApp's decision-making process with the required information.

The xApps lifecycle progresses through three key phases: 
\begin{itemize}
    \item \textbf{Offline Training}: The objective of this phase is to train the DDQN model to learn a policy that optimises UAV mobility and resource allocation based on the defined reward function. This training, based on the O-RAN Alliance specification, will be conducted within the Non-RT RIC, where the operations performed there will not impact network performance. Training data is collected to represent the complex environment and network conditions. The Non-RT RIC acts as a data hub, gathering information from various sources relevant to the network. This could include historical network KPMs, such as UE reports, handover events, and potential network SINR, policy, and configuration information, as well as external data feeds, including the UTM. All this information will be used during the training of our DDQN model, including both individual path-specific models and the global transfer learning model through federated averaging. The training algorithm iteratively updates the weights of the DDQN network to maximise the expected reward. A trained DDQN model represented by the weights of its neural network is the output of this phase.

    \item \textbf{xApp Deployment and Initialisation}: The trained DDQN model is packaged as an xApp container and deployed within the near-RT RIC using the O-RAN xApp management capabilities via the A1 interface. The xApp initialises its internal state, which could include loading the trained DDQN model (either path-specific or global transfer learning model), subscribing to necessary RAN functions through the E2 interface, and setting up communication with EI to receive the updated UAV flight path. The initial state of the xApp, as described in the DDQN model section, is based on the UAV's current location, SINR measurements and the predetermined flight path.

    \item \textbf{Real-time Operation and Control}: The xApp continuously receives data from the corresponding entities to perform the required decisions within the Near-RT RIC's sub-second control loop capabilities. Real-time UAV location updates are received from the UTM as EI, and the xApp subscribes to relevant RAN functions through the E2 interface to obtain real-time measurements from the O-RAN nodes (CU/DU). This includes SINR measurements from each base station associated with this near-RT RIC within range of the UAV, as well as other necessary KPMs, such as network load and channel quality information.
\end{itemize}

The Non-RT RIC serves as a central hub for receiving enrichment information, including the UAV's real-time location from the UTM system. This location data is then transmitted to the Near-RT RIC via the A1 interface, packaged as an A1 EI message containing the UAV's 3D coordinates. Simultaneously, the DUs monitor the signal quality of the UAV's connection, continuously reporting SINR measurements to the Near-RT RIC through the E2 interface. These measurements are sent as E2SM KPM messages.

Once deployed as a container image within the Near-RT RIC, the xApp subscribes to two critical RAN functions through the E2 interface. The first is KPM Reporting (E2SM KPM), which enables the xApp to receive SINR measurements from the DUs. Secondly, the xApp subscribes to RAN Control (E2SM RC), which, if necessary, allows it to initiate handovers or modify resource allocation settings for the UAV. The xApp continuously collects both the UAV location data from the A1 interface and the SINR measurements from the E2 interface, processing this information through the E2 control loop with latencies well-suited for UAV mobility management requirements. These inputs are processed by the DDQN model, which updates its internal state based on this information. The DDQN model, guided by its learned policy and optimised for real-time execution, then decides on the optimal base station for the UAV based on anticipated future signal quality and the UAV's flight path.

The xApp may also adjust bandwidth requirements or proactively initiate handovers if needed by transmitting control actions to the DU via the E2 interface. These actions are encapsulated in E2SM RC messages and can include handover commands, resource requests, or modifications to the UAV's connection settings. The proactive nature of the approach, enabled by predetermined flight path information, reduces sensitivity to control loop latencies by anticipating handover requirements before signal degradation occurs.

To ensure successful deployment and operation, the xApp descriptor, containing its metadata and dependencies, must be carefully configured for the Near-RT RIC environment. Security protocols, including authentication and encryption, are crucial for safeguarding sensitive information, such as the UAV's location and control commands. Furthermore, the performance of the DDQN model and its execution within the Near-RT RIC must be optimised for real-time decision-making, given the dynamic nature of UAV flight. This complex interplay of data collection, intelligent decision-making, and control actions exemplifies how our xApp, when integrated within the O-RAN architecture, can provide a robust solution for managing UAV connectivity and resource allocation in next-generation cellular networks.

\section{Results and Evaluation}\label{ch3_results}
This section presents a comprehensive evaluation of the proposed DDQN-based UAV mobility management framework within O-RAN architectures. The evaluation encompasses multiple critical aspects: the selection and justification of RL methods, the effectiveness of transfer learning across diverse flight scenarios, and a performance comparison against established baseline handover schemes. The analysis demonstrates that the DDQN approach achieves the most favourable performance trade-off compared to both alternative RL methods and traditional handover mechanisms, as validated through extensive simulation across different, distinct flight paths.

\subsection{Simulation Environment}
The simulation environment emulates a realistic UAV operational scenario in a cellular-connected 5G network. The network consists of five BSs placed within an area $2000 \times 2000 \, \text{m}^2$, each equipped with omnidirectional antennas operating at a carrier frequency of 2.1 GHz. All BSs operate at the same carrier frequency, which represents a conservative worst-case interference assumption — in practice, multi-frequency or multi-carrier deployments would reduce inter-cell interference, potentially further improving handover performance. Extending the framework to heterogeneous frequency deployments represents a natural direction for future investigation. BS coordinates are determined to represent a practical urban scenario. The UAV operates at a maximum altitude of 30 meters, with flight paths that incorporate variable altitudes based on urban topology and obstacle avoidance requirements. This includes the takeoff and landing phases, which involve gradually changing altitudes, as well as cruise phases at different operational heights determined by building density and collision avoidance constraints.

\begin{figure*}
    \centering
    \includegraphics[width=\textwidth]{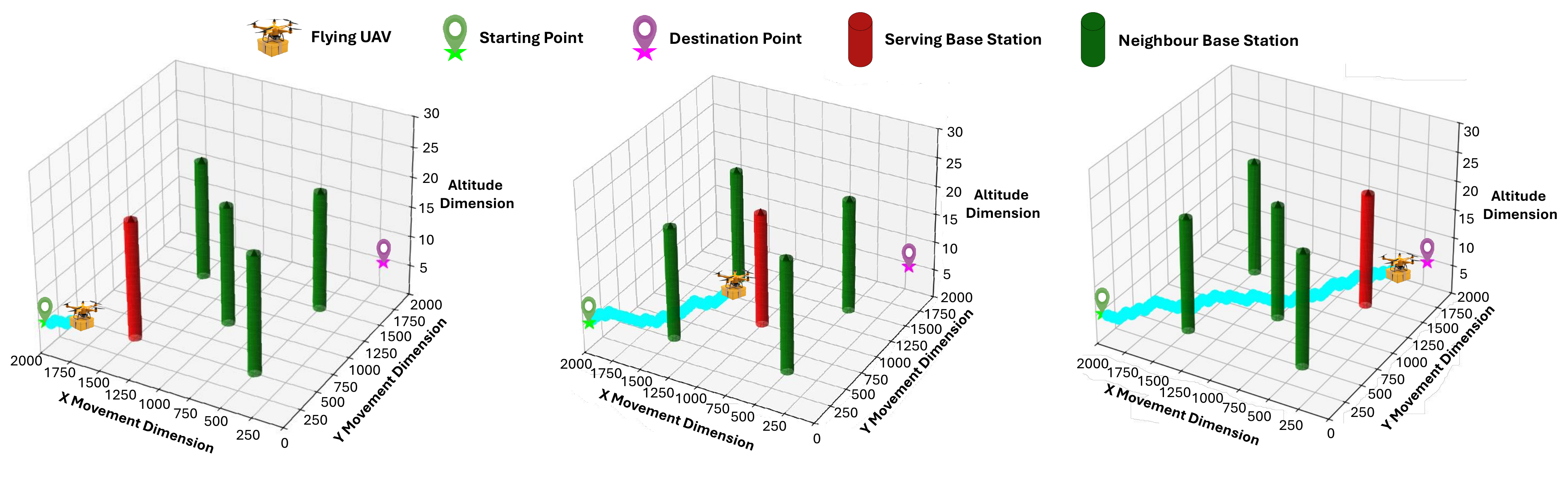}
    \caption{UAV Operational Environment: A 3D visualisation of the UAV's operational environment, including serving and neighbour BS locations and the UAV trajectory}
    \label{fig:uav_environment}
\end{figure*}

Ten distinct flight paths were generated to cover a range of operational conditions, including variations in trajectory patterns, altitude profiles, signal strength environments, and scenarios involving base station coverage overlap. These paths represent different types of UAV missions, including surveillance, delivery, and inspection operations, across urban conditions. The simulation environment addresses realistic operational constraints while providing comprehensive coverage of UAV-network interaction patterns. The simulation environment is visualised in Fig.~\ref{fig:uav_environment}, which represents the 3D operational space of the UAV, including the BS and a representative UAV trajectory..

Each BS transmits with a power of $45 \, \text{dBm}$, and the total noise power is set to $-100 \, \text{dBm}$, corresponding to a noise power spectral density of $-174 \, \text{dBm/Hz}$ over an effective channel bandwidth of approximately $50 \, \text{MHz}$. The UAV travels at a speed of $10 \, \text{m/s}$, and the simulation operates with a discrete timestep of $\Delta t = 0.1 \, \text{s}$, resulting in a spatial resolution of $1 \, \text{m}$ per timestep. This fine-grained temporal discretisation ensures accurate trajectory modelling and enables responsive handover decision-making aligned with the Near-RT RIC's sub-second control loop capabilities. The UAV's SINR for handover decisions is calculated under LoS and NLoS channel conditions, per 3GPP recommendations for UAV mobility in 5G networks, and incorporates both horizontal and vertical distances between the UAV and the BS. Each flight path is evaluated over 500 episodes to ensure statistical reliability of performance metrics. Training data is generated entirely through simulation. At each timestep, the UAV advances along its predetermined flight path, and the simulation computes the SINR received from all BSs using the channel model defined in Section~\ref{ch.3_problem_form}, accounting for LoS and NLoS probabilities, path loss, and inter-cell interference. These computed SINR values, together with the UAV position and the serving BS identity, constitute the state observations observed by the DDQN agent during training. No real network measurements or external datasets are used; the agent learns entirely from interactions with the simulated environment across repeated episodes. The network configuration parameters and simulation assumptions are summarised in Table~\ref{tab:network_params}.

\begin{figure*}[t]
    \centering
    \includegraphics[width=\linewidth]{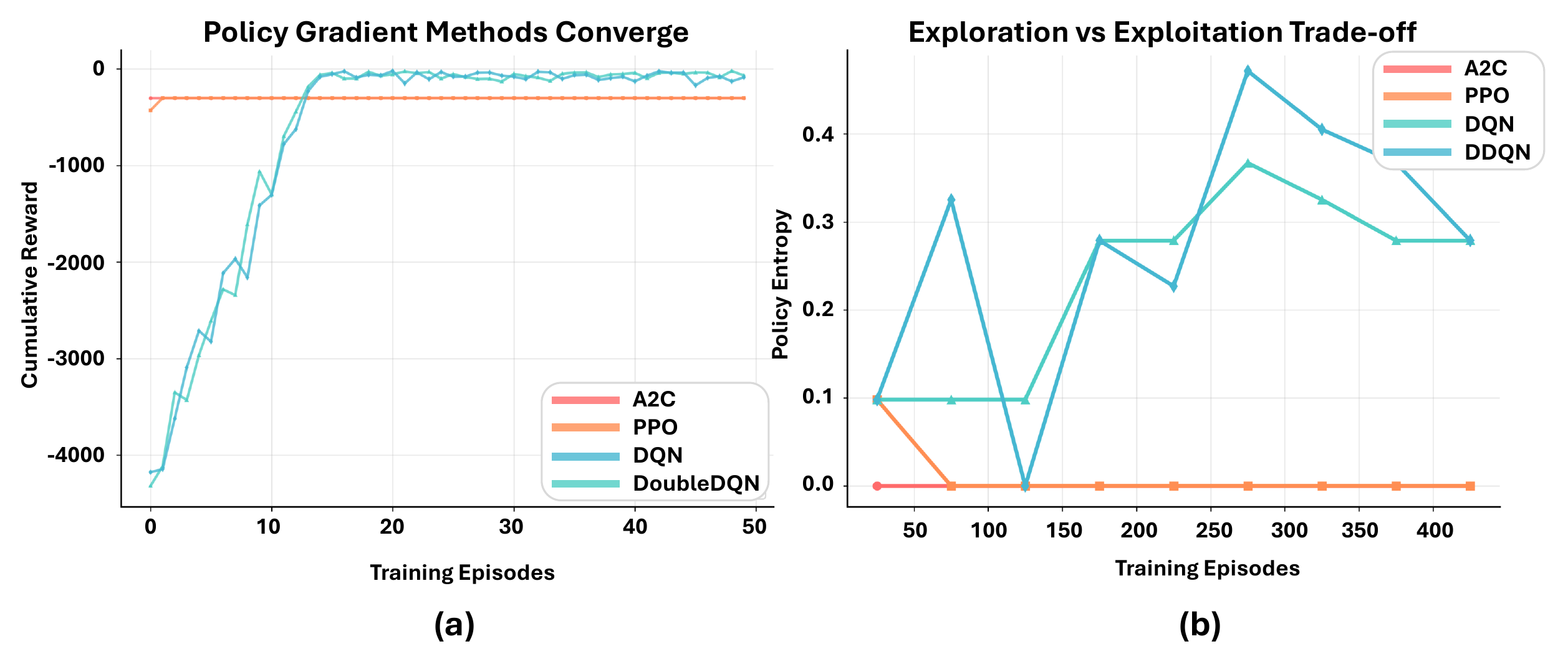}
    \caption{RL Methods Comparison: (a) Training reward convergence, (b) Policy entropy analysis}
    \label{fig:rl_comparison}
\end{figure*}

The following assumptions underpin the simulation:
\begin{itemize}
    \item The UAV maintains continuous connectivity to one BS at any given time, and handovers are executed based on the SINR of candidate BSs and the intelligent learning framework.
    \item Path loss is calculated using separate models for LoS and NLoS conditions, ensuring an accurate representation of the UAV-to-BS channel characteristics based on 3GPP UMa models.
    \item Interference is modelled as the aggregated power from all non-serving BSs, and SINR calculations incorporate this interference to ensure realistic handover dynamics.
    \item Flight paths are predetermined based on mission requirements and obstacle avoidance, representing practical UAV operational scenarios rather than arbitrary trajectories.
    \item The simulation area is bounded, and the BSs are fixed at known locations to facilitate reproducibility of results across different learning methods and baseline comparisons.
\end{itemize}

\begin{table}
    \centering
    \caption{Network Configuration and Simulation Parameters}
    \label{tab:network_params}
    \begin{tabular}{|l|l|}
        \hline
        \textbf{Parameter} & \textbf{Value} \\ \hline
        Simulation Area & $2000 \times 2000 \, \text{m}^2$ \\ \hline
        UAV Max Altitude & $30 \, \text{m}$ \\ \hline
        Number of Flight Paths & 10 \\ \hline
        Episodes per Path & 500 \\ \hline
        Number of BSs & 5 \\ \hline
        BS Transmit Power & $45 \, \text{dBm}$ \\ \hline
        Channel Model & 3GPP UMa \\ \hline
        Noise Power ($N_0$) & $-100 \, \text{dBm}$ \\ \hline
        Carrier Frequency & $2.1 \, \text{GHz}$ \\ \hline
        UAV Speed & $10 \, \text{m/s}$ \\ \hline
        Simulation Timestep ($\Delta t$) & $0.1 \, \text{s}$ \\ \hline
        Spatial Resolution & $1 \, \text{m/step}$ \\ \hline
    \end{tabular}
\end{table}

\subsection{Reinforcement Learning Methods Configuration}

The selection of an appropriate RL algorithm for UAV handover optimisation requires careful consideration of the problem characteristics, including discrete action spaces, temporal decision dependencies, and the need for stable learning in dynamic environments. This subsection presents a comprehensive comparison of different RL approaches and justifies the selection of DDQN as the primary method for the proposed mobility management framework.

Initial experiments evaluated multiple RL paradigms to identify the most suitable approach for UAV handover optimisation. Figure \ref{fig:rl_comparison} demonstrates the fundamental performance differences between value-based methods (DQN, DDQN) and policy gradient approaches (PPO, A2C). The results reveal that policy gradient methods exhibit severe limitations in this application domain, converging to suboptimal policies that avoid handovers entirely regardless of signal quality conditions.

As illustrated in Figure \ref{fig:rl_comparison}(a), both DQN and DDQN achieve effective learning with rewards converging toward higher values near zero. At the same time, PPO and A2C remain stuck at poor performance levels, with rewards ranging from -300 to -500. The policy entropy analysis in Figure \ref{fig:rl_comparison}(b) reveals the underlying cause: policy gradient methods rapidly lose exploration capability, with entropy approaching zero, indicating premature convergence to deterministic policies that maintain current connections regardless of signal conditions. This behaviour arises from the immediate penalty structure of handover decisions, where policy gradient optimisation prioritises immediate reward maximisation and fails to explore actions with short-term costs but long-term connectivity benefits.

\begin{figure*}
    \centering
    \includegraphics[width=\linewidth]{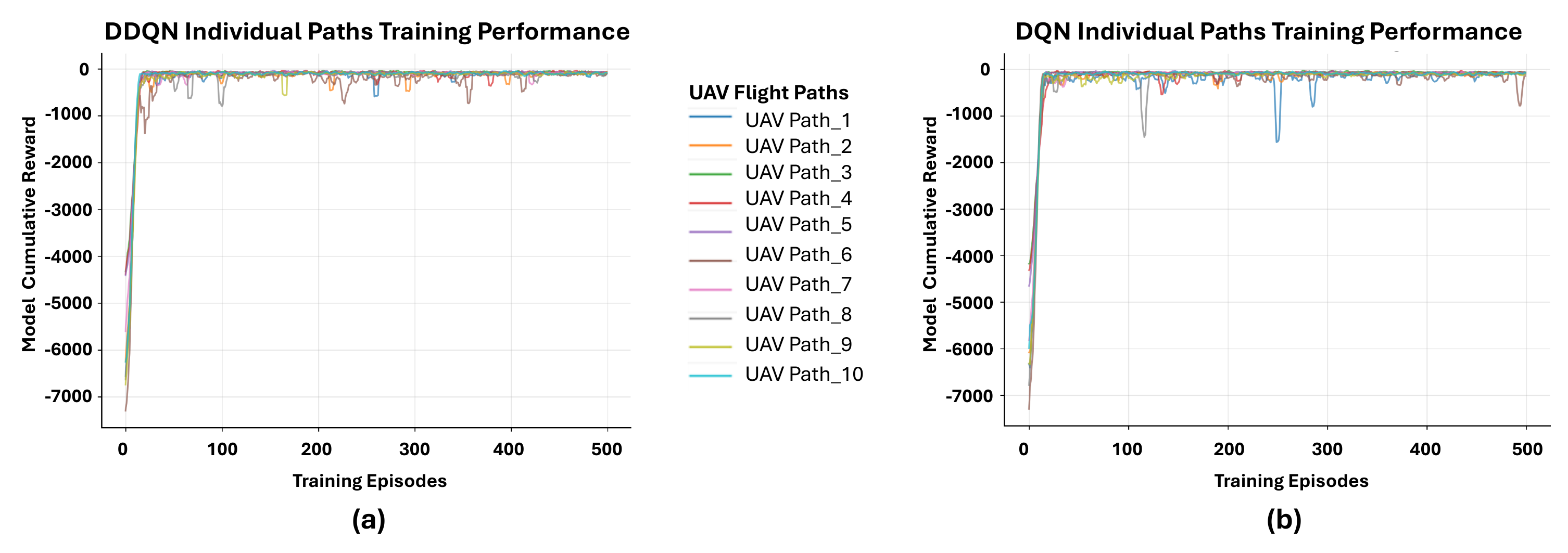}
    \caption{Individual Path Training Performance: (a) DQN training rewards across distinct paths, (b) DDQN training rewards across distinct paths}
    \label{fig:individual_training}
\end{figure*}

Having established the superiority of value-based methods, a detailed comparison between DQN and DDQN reveals the benefits of addressing overestimation bias in Q-value updates. Figure \ref{fig:individual_training} presents the training performance of both algorithms across individual flight paths, demonstrating DDQN's enhanced stability and convergence characteristics.

\begin{table}
\centering
\caption{Double Deep Q-Network Training Hyperparameters}
\label{tab:dqn_parameters}
\begin{tabular}{|p{4cm}|p{3cm}|}
\hline
\textbf{Hyperparameter} & \textbf{Value} \\
\hline
Hidden Layer Size & 128 \\
Learning Rate ($\alpha$) & 0.001 \\
Discount Factor ($\gamma$) & 0.99 \\
$\epsilon$-greedy rate ($\epsilon$) & 1.0 $\rightarrow$ 0.01 \\
Target Network Update Frequency & 1000 steps \\
Replay Buffer Size & 10000 \\
Batch Size & 64 \\
\hline
\end{tabular}
\end{table}

The individual path training results in Figure \ref{fig:individual_training} show that both algorithms achieve effective learning. Still, DDQN demonstrates more consistent convergence patterns and reduced variance in final performance across different flight scenarios. DDQN's separation of action selection and evaluation phases leads to more stable Q-value estimates, which is particularly beneficial in dynamic UAV environments where signal conditions vary rapidly.

\begin{figure}
    \centering
    \includegraphics[width=\linewidth]{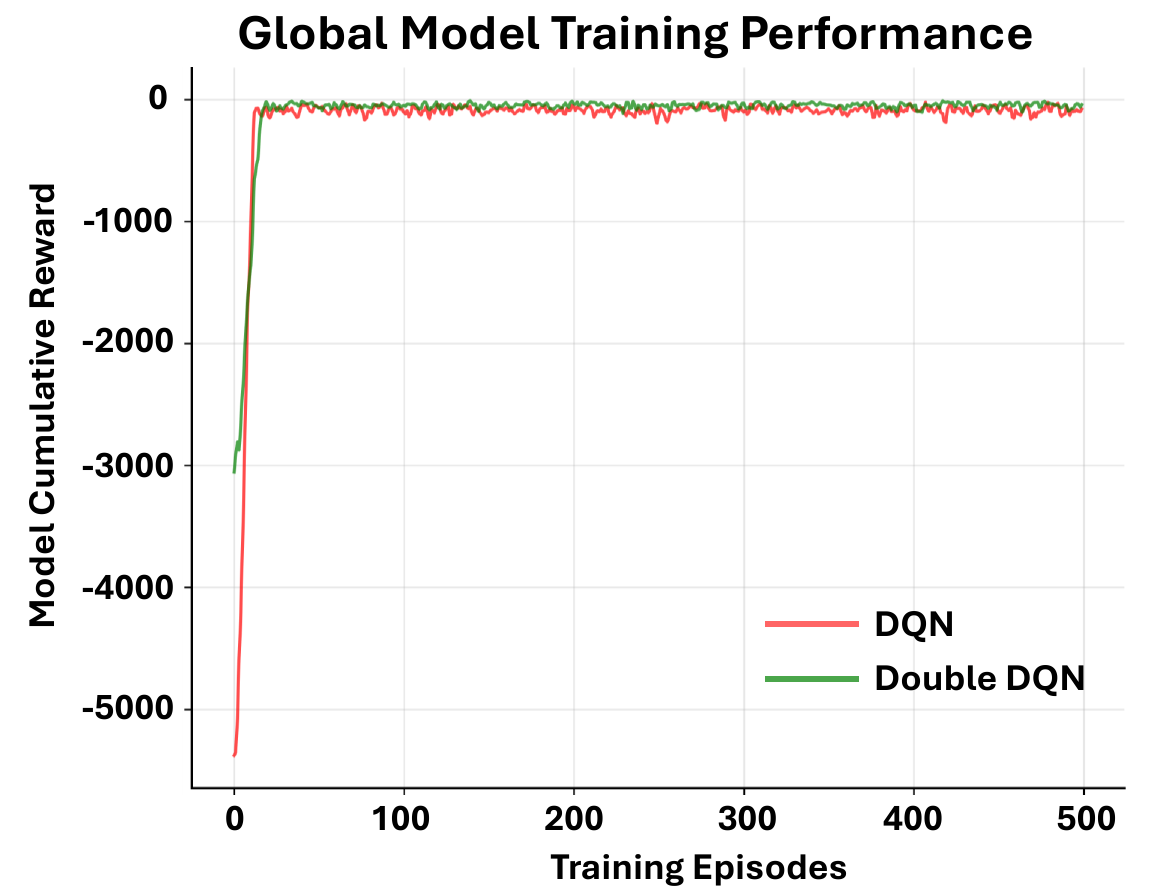}
    \caption{Global Model Training Performance: DDQN and DQN global model fine-tuning on new flight path}
    \label{fig:global_training}
\end{figure}

The effectiveness of the weight averaging and transfer learning approach is demonstrated by its impact on global model training performance. Figure \ref{fig:global_training} compares the fine-tuning performance of global models derived from both DQN and DDQN individual models.

The global model training results demonstrate that the DDQN-based global model achieves faster convergence and better final performance during fine-tuning on new flight paths. This superior adaptation capability validates the choice of DDQN as the foundation for the transfer learning framework, where stable Q-value estimates prove crucial for effective knowledge transfer across diverse flight scenarios.

The DDQN configuration employs carefully tuned hyperparameters optimised for UAV mobility scenarios: learning rate $\alpha$ = 0.001, discount factor $\gamma$ = 0.99, exploration rate $\epsilon$ decreasing from 1.0 to 0.01, target network update frequency of 1000 steps, replay buffer size of 10,000 experiences, and batch size of 64. The network architecture consists of three fully connected layers, each with 128 hidden units, utilising ReLU activation functions for the hidden layers and a linear activation function for the output layer. Table~\ref{tab:dqn_parameters} summarises the key hyperparameters used in the DDQN training process.

Based on this comprehensive analysis, DDQN emerges as the optimal choice for UAV handover optimisation, combining the benefits of value-based learning with enhanced stability through a double network architecture. The superior performance in both individual training and global model transfer learning validates this selection for the proposed O-RAN mobility management framework.

\subsection{Baseline Handover Schemes}

To evaluate the effectiveness of the proposed DDQN-based mobility management approach, a comprehensive comparison is conducted against three established handover strategies representing different levels of complexity and decision-making sophistication. These baseline methods provide benchmarks for assessing the performance gains achieved through intelligent learning-based optimisation.

\textbf{Greedy Handover Scheme:} This scheme represents the most straightforward reactive approach to handover management, where the UAV instantaneously switches to the base station providing the strongest received signal strength. The handover decision is triggered immediately when the SINR from a neighbouring base station exceeds that of the currently connected base station by a predetermined margin $\Psi$:
\begin{equation}
\gamma_i(t) > \gamma_k(t) + \Psi, \quad \forall i \in \mathcal{N}_k(t)
\end{equation}
where $\gamma_i(t)$ represents the SINR from base station $i$ at time $t$, $k$ indexes the currently connected base station, and $\mathcal{N}_k(t)$ denotes the set of neighboring base stations. This approach prioritises signal strength maximisation without considering handover frequency or long-term connectivity implications, making it susceptible to ping-pong effects and excessive handover events.

\textbf{Hysteresis-Based Handover Scheme:} To mitigate the ping-pong effects inherent in greedy approaches, this method incorporates temporal stability through hysteresis margin and time-to-trigger mechanisms. A handover is initiated only when the SINR from a neighbouring base station consistently exceeds the current serving base station by a margin $\Psi$ for a duration exceeding the time-to-trigger parameter $\Omega$:
\begin{equation}
\gamma_i(t) > \gamma_k(t) + \Psi, \quad \forall t \in [t - \Omega, t], \quad \forall i \in \mathcal{N}_k(t)
\end{equation}
This constraint ensures handover execution only when signal improvement is sustained over the specified time window, reducing unnecessary handovers caused by temporary signal fluctuations. The performance of this scheme depends critically on the appropriate selection of hysteresis margin $\Psi$ and time-to-trigger $\Omega$ parameters.

\begin{figure*}
    \centering
    \includegraphics[width=\linewidth]{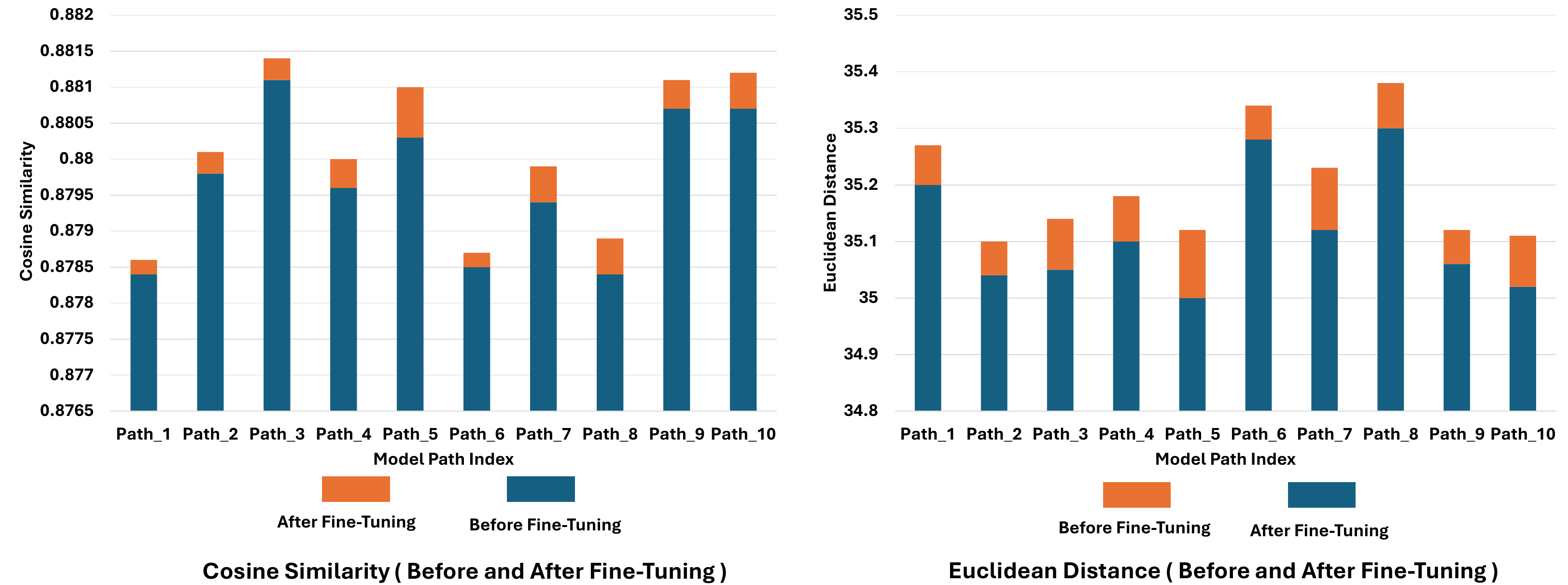}
    \caption{Transfer Learning Effectiveness Analysis: Cosine similarity and Euclidean distance measurements between global model and individual models before and after fine-tuning}
    \label{fig:transfer_learning_analysis}
\end{figure*}

\textbf{Minimum Outage Probability (MOP) Scheme:} This sophisticated approach proactively minimises communication link outages by selecting the base station with the lowest predicted outage probability over a specified prediction horizon $H$. The handover decision is formulated as an optimisation problem:
\begin{equation}
k^* = \arg\min_i P_{\text{out},i}, \quad i \in \{1, 2, \ldots, M\}
\end{equation}
where $P_{\text{out},i}$ represents the predicted outage probability for base station $i$, and $M$ denotes the total number of base stations. The outage probability prediction considers anticipated signal degradation patterns based on UAV trajectory and historical network performance. While this predictive approach offers enhanced robustness against connectivity disruptions, its effectiveness depends on accurate estimation of outage probabilities, which can be challenging in highly dynamic UAV environments.

All baseline schemes are implemented with carefully tuned parameters to ensure fair comparison: greedy margin $\Psi = 3$ dB, hysteresis margin $\Psi = 3$ dB with time-to-trigger $\Omega = 2$ seconds, and MOP prediction horizon $H = 5$ time steps. These parameter values are selected based on established practices in cellular handover management and preliminary optimisation to avoid unfavourable bias against baseline methods. The comparative evaluation against these benchmarks demonstrates the advantages of intelligent learning-based approaches that can adapt to dynamic network conditions and UAV mobility patterns while balancing multiple optimisation objectives simultaneously.

\subsection{Performance Results}

The effectiveness of the proposed transfer learning framework is demonstrated through a comprehensive analysis of knowledge consolidation and generalisation capabilities across diverse flight scenarios. Figure \ref{fig:transfer_learning_analysis} presents the weight similarity analysis before and after fine-tuning, providing empirical evidence for the successful transfer of knowledge between the global model and individual path-specific models.

\begin{table}
    \centering
    \caption{Performance Summary: Mean and Standard Deviation of Handover Events and Outage Probability Across All Methods}
    \label{tab:performance_summary}
    \begin{tabular}{|l|c|c|c|c|}
        \hline
        \textbf{Method} & \multicolumn{2}{c|}{\textbf{Handover Events}} & \multicolumn{2}{c|}{\textbf{Outage Probability (\%)}} \\
        \cline{2-5}
        & \textbf{Mean} & \textbf{Std. Dev.} & \textbf{Mean} & \textbf{Std. Dev.} \\
        \hline
        DDQN & 2.45 & 0.73 & 0.048 & 0.021 \\
        DQN & 2.69 & 0.82 & 0.052 & 0.023 \\
        Greedy & 5.40 & 1.12 & 0.021 & 0.011 \\
        Hysteresis & 4.30 & 0.61 & 0.089 & 0.047 \\
        MOP & 4.00 & 0.87 & 0.041 & 0.019 \\
        \hline
    \end{tabular}
\end{table}

The cosine similarity analysis reveals substantial similarity among individual model weights before fine-tuning, with values ranging from 0.878 to 0.881, indicating that models trained on different flight paths learn comparable handover decision patterns. After fine-tuning the global model on previously unseen flight scenarios, similarity scores improve to 0.8785-0.8812. At the same time, Euclidean distances decrease from approximately 35.3 to 35.0, demonstrating that the transfer learning process successfully enhances generalisation while preserving learned knowledge. This empirical evidence directly addresses concerns about the theoretical foundation of the weight averaging approach, showing that common handover optimisation patterns do indeed transfer across diverse flight scenarios.

A comparative performance analysis across all evaluated methods demonstrates the clear superiority of the DDQN-based approach over both the DQN and traditional baseline schemes. Table \ref{tab:performance_summary} presents a comprehensive statistical analysis, including mean performance and standard deviation across 500 evaluation episodes for each method. The outage probability values represent the empirical outage rate $\hat{P}_{\text{out}}$, computed as the fraction of time steps during which the received SINR falls below the minimum threshold $\gamma_{\text{th}}$, averaged across the evaluation episodes.

The statistical analysis reveals that DDQN achieves the lowest handover frequency (2.45 events on average) with a 54.6\% reduction compared to the greedy scheme. While the greedy scheme achieves the lowest outage probability (0.021\%), this comes at the cost of more than double the handover events (5.40), which in practice translates to substantially higher signalling overhead, increased risk of radio link failures, and greater service disruption during handover execution. In contrast, DDQN maintains outage probability at a practically negligible level (0.048\%), representing an absolute difference of only 0.027 percentage points relative to the greedy baseline---a marginal increase that is operationally insignificant for most UAV mission profiles. Furthermore, DDQN outperforms the hysteresis scheme on both metrics simultaneously, achieving fewer handovers (2.45 vs 4.30) and substantially lower outage (0.048\% vs 0.089\%). Compared to MOP, DDQN achieves 38.8\% fewer handovers with comparable outage levels (0.048\% vs 0.041\%). The improvement over DQN (8.9\% reduction in handover events) validates the benefits of addressing overestimation bias through the double network architecture.

\begin{figure}[t]
    \centering
    \includegraphics[width=\linewidth]{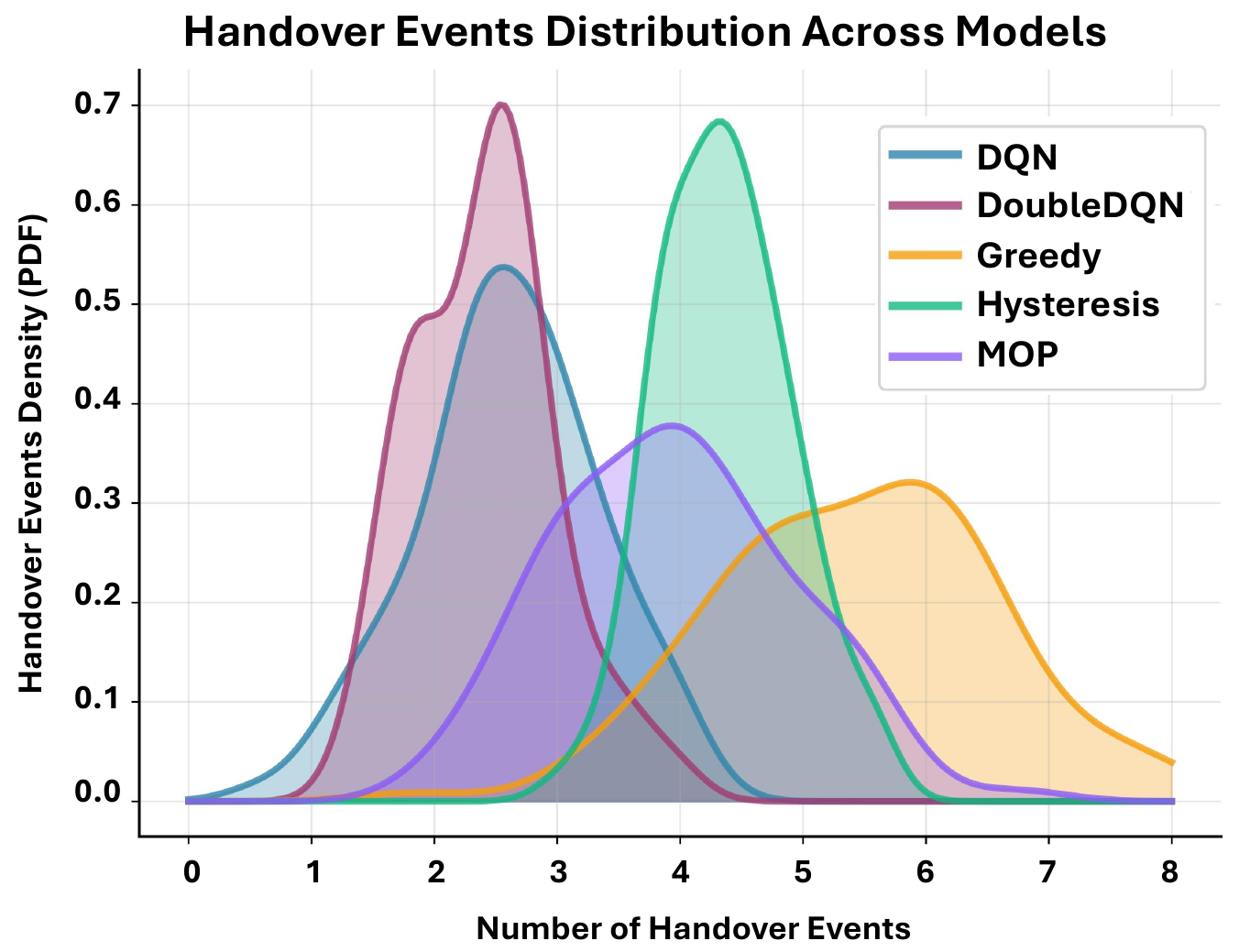}
    \caption{Handover Events Distribution Across Models}
    \label{fig:handover_distribution}
\end{figure}

Figure \ref{fig:handover_distribution} illustrates the distribution of handover events across all evaluated methods, revealing the concentrated performance of learning-based approaches compared to the wider distribution exhibited by traditional baseline schemes.

The handover distribution analysis shows that DDQN consistently maintains low handover frequencies with minimal variance, resulting in a tight distribution centred around 2-3 handover events per episode. In contrast, baseline methods exhibit significantly higher variance, with the greedy approach showing particularly poor consistency and frequent instances of excessive handover activity.

\begin{figure}[t]
    \centering
    \includegraphics[width=\linewidth]{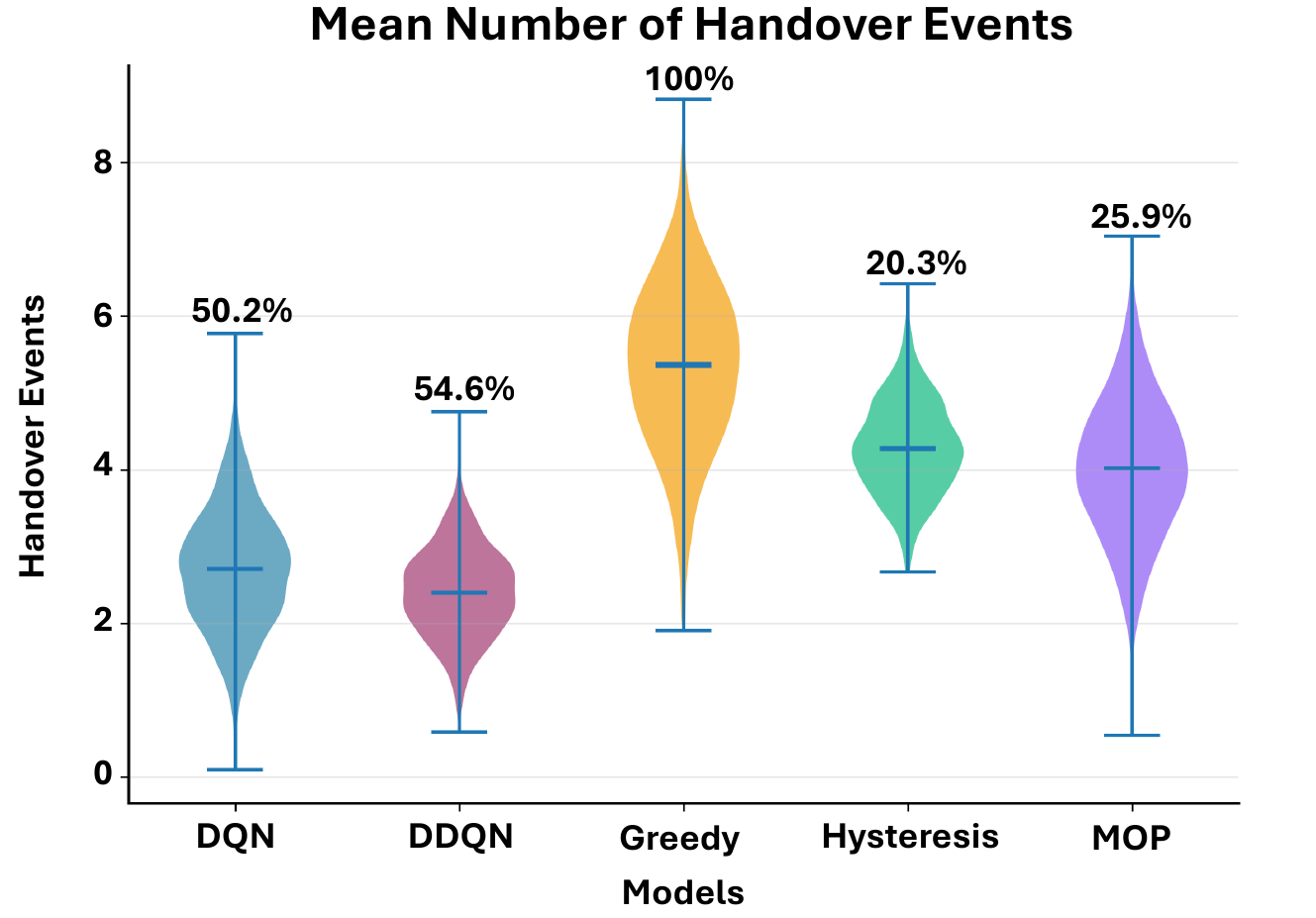}
    \caption{Handover Performance Comparison: Mean handover events with percentage reduction relative to greedy baseline}
    \label{fig:handover_performance}
\end{figure}

A detailed comparison of handover performance across methods is presented in Figure \ref{fig:handover_performance}, which quantifies the percentage reduction in handover events achieved by each approach relative to the worst-performing baseline by presenting the mean of the handover events in each model.

\begin{figure}[t]
    \centering
    \includegraphics[width=\linewidth]{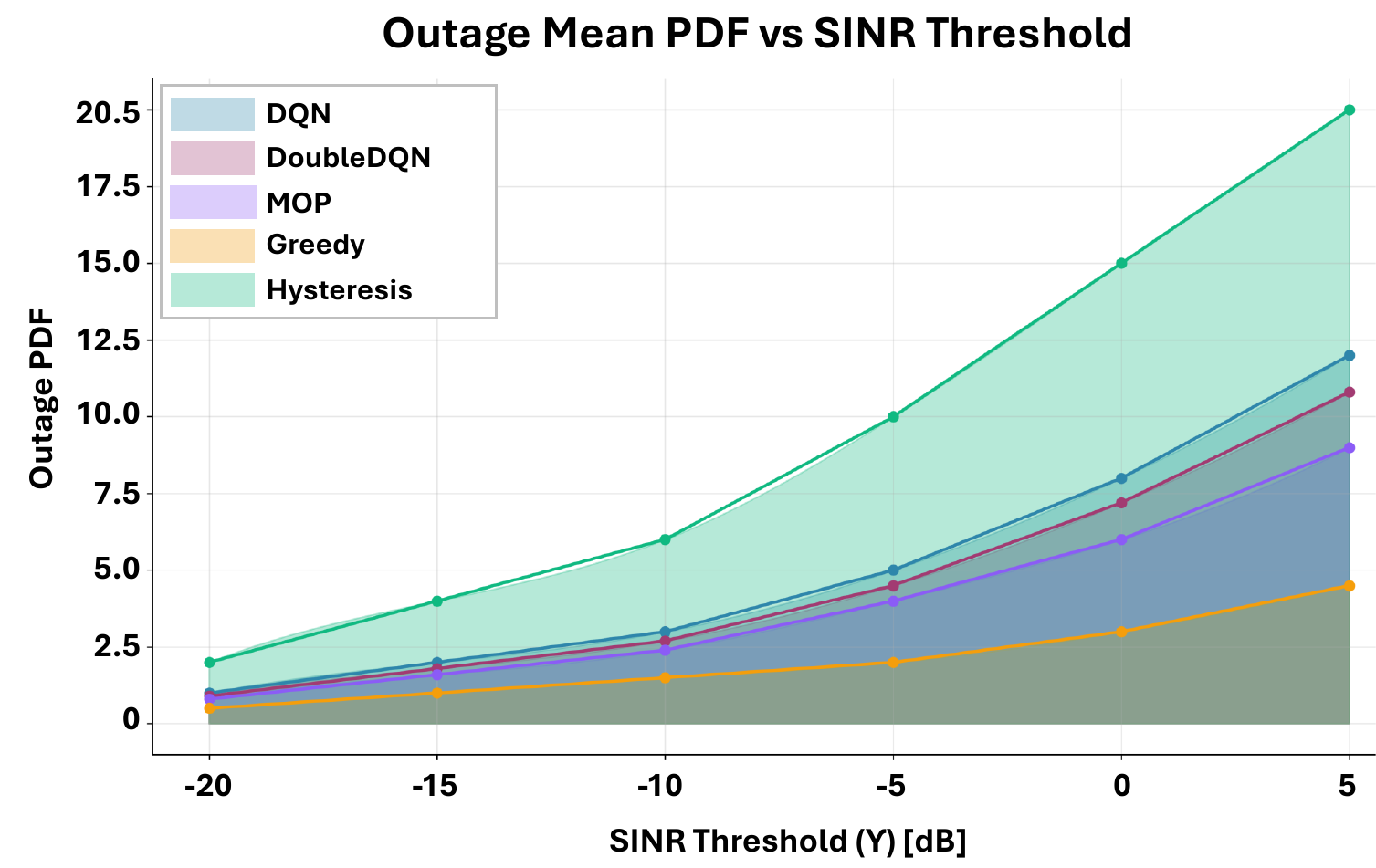}
    \caption{Outage Performance Sensitivity Analysis: Mean outage probability versus SINR threshold}
    \label{fig:outage_sensitivity}
\end{figure}

The robustness of the proposed approach across varying network conditions is evaluated through a sensitivity analysis of SINR thresholds. Figure \ref{fig:outage_sensitivity} presents outage probability performance as a function of SINR threshold values, demonstrating how different methods respond to changing signal quality requirements.

The sensitivity analysis confirms the trade-off characteristics of each method across varying SINR thresholds. The greedy approach achieves the lowest outage probability under stringent SINR requirements, which is expected since it always selects the strongest available signal regardless of handover cost. However, as demonstrated in Table~\ref{tab:performance_summary}, this outage advantage comes at a disproportionate cost in handover frequency. DDQN maintains competitive outage performance across all evaluated thresholds while consistently delivering substantial handover reduction, demonstrating the robustness of the learned policy in balancing both objectives. The hysteresis approach exhibits the highest outage rates due to its conservative handover policy, which delays necessary handovers and results in prolonged connections to degrading base stations. The MOP scheme demonstrates intermediate performance on both metrics, offering a reasonable but non-adaptive balance.

The comprehensive evaluation demonstrates that the proposed DDQN-based mobility management framework achieves the most favourable trade-off between handover frequency minimisation and connectivity reliability among all evaluated methods. By substantially reducing handover events while maintaining outage probability at operationally negligible levels, the framework addresses the practical requirements of UAV missions where excessive handovers pose greater risks to mission continuity than marginal differences in outage rate. The transfer learning approach proves effective for generalisation across diverse flight scenarios, addressing scalability concerns for practical O-RAN deployment. These results validate the selection of DDQN as a robust foundation for intelligent UAV mobility management in next-generation cellular networks.

\section{Conclusion}

This work presents an intelligent UAV mobility management framework for O-RAN architectures, utilising a DDQN RL algorithm enhanced with transfer learning. The proposed system enables proactive handover optimisation for UAVs operating along predetermined flight trajectories, transforming reactive connectivity management into predictive resource allocation.

The evaluation demonstrates that DDQN achieves the most favourable balance between handover reduction and connectivity reliability among all evaluated methods, reducing handover frequency by up to 54.6\% compared to greedy approaches while maintaining outage probability at practically negligible levels. Although the greedy baseline achieves marginally lower outage rates, this comes at a disproportionate cost in handover frequency, which in operational UAV deployments translates to increased signalling overhead and elevated risk of radio link failures. The framework successfully addresses the limitations of policy gradient methods (PPO, A2C) that failed due to premature convergence, while outperforming standard DQN through enhanced stability provided by a double network architecture.

A key contribution is the successful implementation of transfer learning through centralised weight averaging, consolidating knowledge from multiple flight scenarios into a robust global model. Empirical analysis provides concrete evidence that handover decision patterns transfer effectively across diverse operational scenarios, enabling deployment without the need for extensive retraining in new environments.

The integration within O-RAN Near-RT RIC leverages sub-second control capabilities suitable for UAV handover requirements, with proactive optimisation mitigating latency sensitivity. The results validate the effectiveness of RL for UAV mobility management in next-generation cellular networks, demonstrating that intelligent multi-objective optimisation can achieve a superior trade-off between handover reduction and connectivity stability compared to both reactive and predictive baseline schemes.

Future work will explore multi-UAV scenarios with coordinated handover management, priority-aware reward shaping for mission-critical UAV operations, and validation in larger network deployments to assess scalability. Regarding computational complexity, the DDQN framework incurs a one-time offline training cost, while inference at deployment is computationally negligible as discussed in Section~\ref{ch3_results}. The 54.6\% reduction in handover events over the greedy baseline translates directly to reduced signalling overhead and fewer radio link failures in operational deployments, justifying the added intelligence over simpler reactive schemes. Real-world O-RAN testbed validation remains a key next step.

\section*{Acknowledgment}
This research was funded by the Engineering and Physical Sciences Research Council (EPSRC) grants EP/Y037421/1 and EP/X040518/1 under the Communications Hub for Empowering Distributed clouD computing Applications and Research (CHEDDAR) project.

\bibliographystyle{IEEEtran}
\bibliography{reference}


\begin{IEEEbiography}[{\includegraphics[width=1in,height=1.15in,clip]{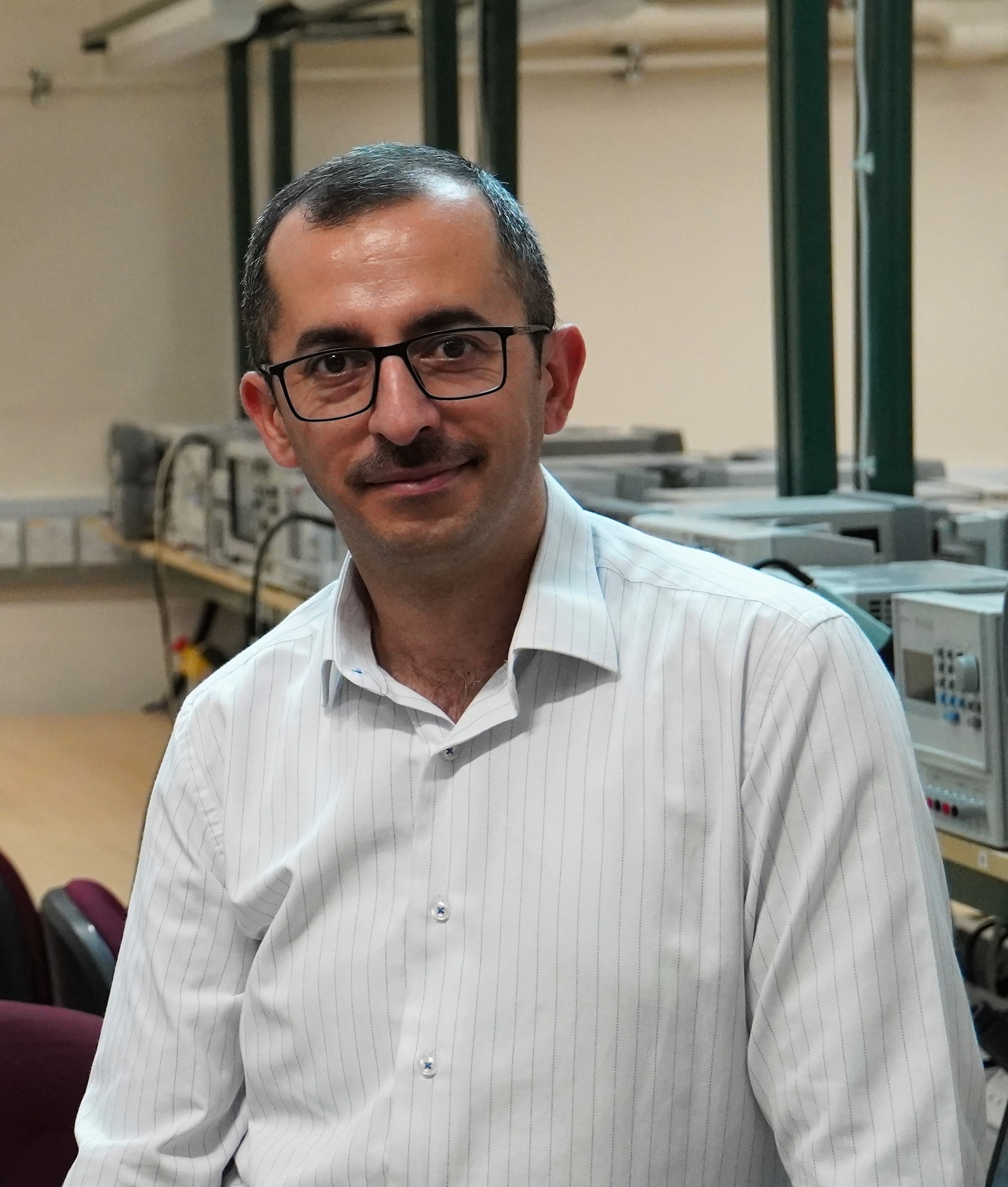}}]{Mohammed M. H. Qazzaz}\hspace{0.075 cm} is an accomplished academic and researcher who earned his BSc and MSc degrees in Electronics \& Communication Engineering from Mosul University, Iraq, in 2010 and 2013, respectively. He is currently pursuing a PhD degree at the University of Leeds, UK, with a research focus on Wireless Communications, UAV Communications, Artificial Intelligence / Machine Learning, and Radio Access Networks (RANs). Prior to his doctoral studies, he served as an assistant lecturer at the Communication Engineering School, Ninevah University, Iraq, from January 2017 to January 2022, where he shared his knowledge and expertise. He also brings seven years of practical experience from the telecommunications industry, working in various roles with local and international vendors and operators.
\end{IEEEbiography}

\begin{IEEEbiography}[{\includegraphics[width=1.1in,height=2in,clip,keepaspectratio]{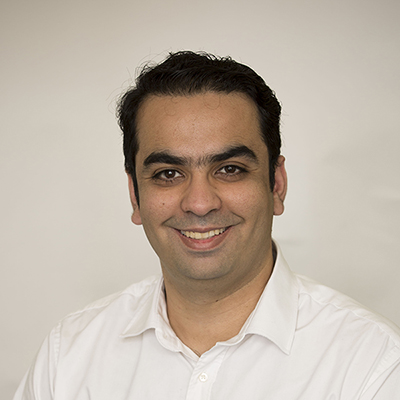}}]{Syed Ali Raza Zaidi } (Member, IEEE)  is an Associate Professor at the University of Leeds in the area of Communication and Sensing for Robotics and Autonomous Systems. He co-leads the UK’s Department for Science, Innovation and Technology (DSIT) and UKRI-funded Future Communications Hub for Empowering Distributed Cloud Computing Applications and Research (CHEDDAR), which has received £16 million in research funding. He leads the Emergent Compute Pillar within the CHEDDAR work programme, as well as DSIT- and AISI-funded initiatives on agentic AI for cloud-native telecommunications.
Earlier, 2013 to 2015, he was associated with the SPCOM research group, working on a US ARL-funded project in Network Science. From 2011 to 2013, he was a research associate at the International University of Rabat. He was a visiting research scientist at Qatar Innovations and Mobility Centre from October to December 2013, where he worked on the QNRF-funded project QSON. He completed his doctoral degree at the School of Electronic and Electrical Engineering, where he was awarded the G. W. and F. W. Carter Prize for best thesis and best research paper. He has published over 90 papers in leading IEEE conferences and journals. From 2014 to 2015, he served as an editor of IEEE Communication Letters and as the lead guest editor for the IET Signal Processing Journal’s Special Issue on Signal Processing for Large-Scale 5G Wireless Networks. He has served as lead editor for the IEEE Communications Magazine Feature Topic on Communication Technologies for Robotics and Autonomous Systems and for the IEEE Journal on Selected Areas in Communications (JSAC) Special Issue on Design and Analysis of Communication Interfaces for Industry 4.0. He is an editor for the IET Access, Fronthaul, and Backhaul book series, and is currently an Associate Technical Editor for IEEE Communications Magazine. He is the Industrial Sponsorship and Programme Chair for ICC 2026. He has been awarded grants from COST IC0902, the Royal Academy of Engineering, EPSRC, Horizon Europe, and DAAD (totalling approximately £12 million) to promote his research. He has been an invited keynote speaker and panellist at various leading international conferences and workshops. His current research interests include Generative AI for cloud-native telecommunications and the modelling, analysis, and design of large-scale connected intelligent systems. \end{IEEEbiography}

\begin{IEEEbiography}[{\includegraphics[width=1.05in,height=1.2in,clip]{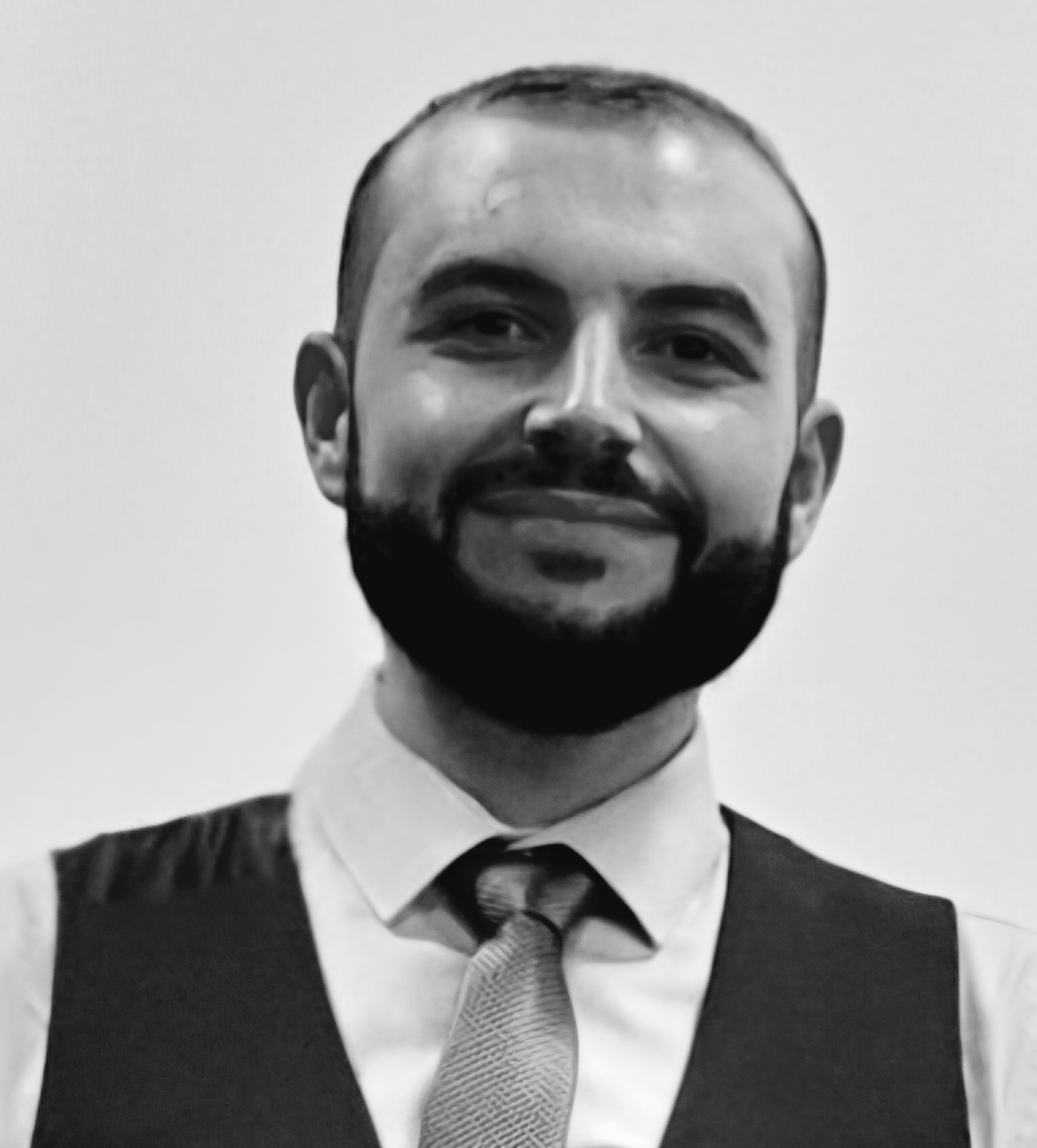}}]{Aubida A. Al-Hameed}\hspace{0.075 cm} received his B.Sc. (2010) and M.Sc. (2013) degrees in electrical engineering from the University of Mosul, Iraq. He obtained his Ph.D. in electronic and electrical engineering from the University of Leeds, U.K., in 2019. Currently a lecturer at the University of Ninevah, his research focuses on optical wireless communications, AI-enabled networks, and machine learning for 5G/6G systems.\end{IEEEbiography}

\begin{IEEEbiography}[{\includegraphics[width=1in,height=1.15in,clip]{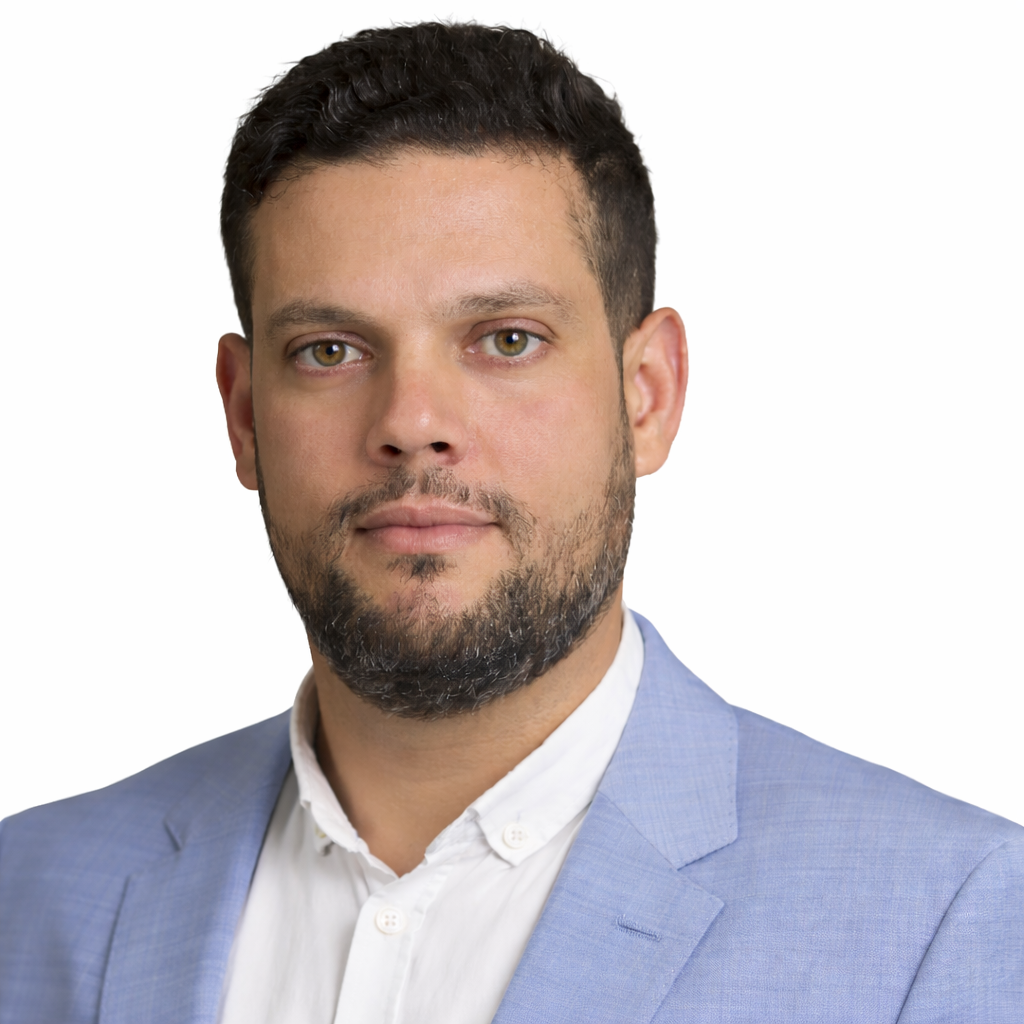}}]{Abdelaziz Salama}\hspace{0.075 cm} received the B.Sc. degree in Communication Engineering from Tripoli University, Tripoli, Libya, in 2009, the M.Sc. degree in Communication, Control, and Digital Signal Processing from the University of Strathclyde, Glasgow, U.K., in 2017, and the Ph.D. degree in Federated Learning and Edge Intelligence for Wireless Mesh Networks from the University of Leeds, Leeds, U.K., in 2024. He is currently a Research Fellow with the University of Leeds, working on O-RAN, Agentic AI, and 6G network architectures within the CHEDDAR Hub project.
His research interests include Open RAN (O-RAN), federated learning, edge intelligence, autonomous and AI-driven wireless networks, cloud-native RAN systems, and Agentic AI for 6G communications. He has contributed to the design and deployment of xApps, Near-RT RIC platforms, and AI-enabled optimisation frameworks for autonomous network management. He is an active contributor to O-RAN and AI-RAN research activities and has presented demonstrations and research work at major international venues, including IEEE conferences and the Mobile World Congress (MWC). Prior to academia, he gained extensive industry experience in telecommunications, satellite communications, IP networking, and mission-critical communication systems across both industrial and international organisations.
\end{IEEEbiography}

\begin{IEEEbiography}[{\includegraphics[width=1.05in,height=1.2in,clip]{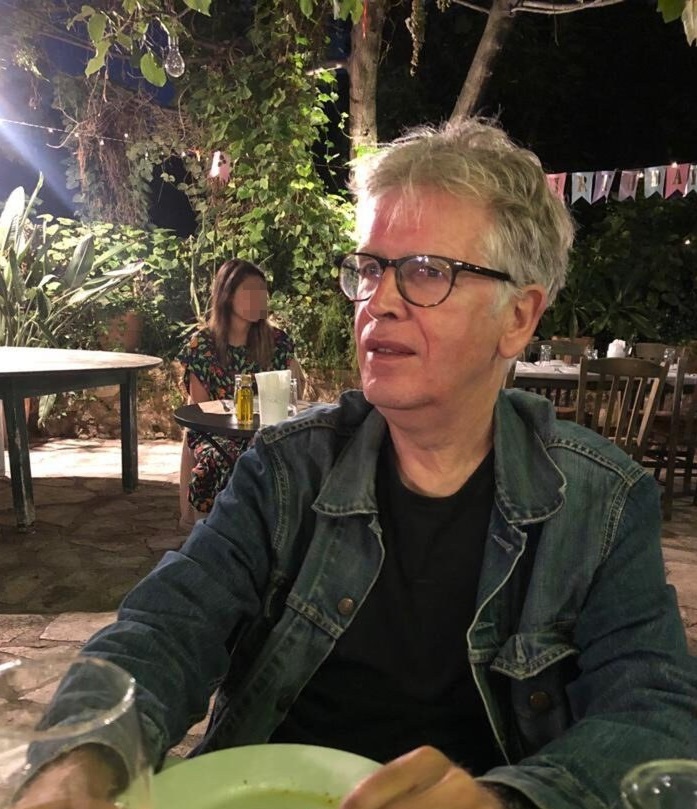}}]{Des McLernon}\hspace{0.075 cm} received both his B.Sc in electronic and electrical engineering and his MSc in electronics from the Queen's University of Belfast, N. Ireland. After working on radar systems research with Ferranti Ltd in Edinburgh, Scotland, he then joined Imperial College, University of London, UK, where he took his PhD in signal processing. He is currently a Reader in Signal Processing at the University of Leeds, UK. His research interests are broadly within the domain of signal processing for wireless communications, in which field he has around 360 research publications and also supervised over 50 PhD students. Finally, in the little spare time that remains, he plays jazz piano in restaurants and bars.\end{IEEEbiography}

\vfill\pagebreak

\end{document}